\documentclass[prb,aps,showpacs,twocolumn,amsmath,amssymb,floatfix,superscriptaddress]{revtex4-2}

\usepackage{hyperref}
\usepackage{graphicx}
\usepackage{bm}
\usepackage{dsfont}
\usepackage{xcolor}
\usepackage[utf8]{inputenc}
\usepackage{amssymb}
\usepackage[normalem]{ulem}
\usepackage{physics}

\newcommand{\beq}{\begin{equation}}
\newcommand{\eeq}{\end{equation}}
\newcommand{\beqa}{\begin{eqnarray}}
\newcommand{\eeqa}{\end{eqnarray}}

\begin{document}

\title{Phenomenology of Majorana zero modes in full-shell hybrid nanowires}
\author{Carlos Payá}
\affiliation{Instituto de Ciencia de Materiales de Madrid (ICMM), CSIC, Madrid, Spain}
\author{Samuel D. Escribano}
\affiliation{Department of Condensed Matter Physics, Weizmann Institute of Science, Rehovot 7610, Israel}
\author{Andrea Vezzosi}
\affiliation{Dipartimento di Scienze Fisiche, Informatiche e Matematiche, Universit\`a di Modena e Reggio Emilia, Via Campi 213/a, 41125 Modena, Italy}
\author{Fernando Peñaranda}
\affiliation{Donostia International Physics Center, P. Manuel de Lardizabal 4, 20018 Donostia-San Sebastian, Spain}
\author{Ramón Aguado}
\affiliation{Instituto de Ciencia de Materiales de Madrid (ICMM), CSIC, Madrid, Spain}
\author{Pablo San-Jose}
\affiliation{Instituto de Ciencia de Materiales de Madrid (ICMM), CSIC, Madrid, Spain}
\author{Elsa Prada}
\affiliation{Instituto de Ciencia de Materiales de Madrid (ICMM), CSIC, Madrid, Spain}

\date{\today}
 \begin{abstract}
Full-shell nanowires have been proposed as an alternative nanowire design in the search of topological superconductivity and Majorana zero modes (MZMs). They are hybrid nanostructures consisting of a semiconductor core fully covered by a thin superconductor shell and subject to a magnetic flux. Compared to their partial-shell counterparts, full-shell nanowires present some advantages that could help to clearly identify the elusive Majorana quasiparticles, such as the operation at smaller magnetic fields and low or zero semiconductor g-factor, and the expected appearance of MZMs at well-controlled regions of parameter space. In this work we critically examine this proposal, finding a very rich spectral phenomenology that combines the Little-Parks modulation of the parent-gap superconductor with flux, the presence of flux-dispersing Caroli--de Gennes--Matricon (CdGM) analog subgap states, and the emergence of MZMs across finite flux intervals that depend on the transverse wavefunction profile of the charge density in the core section.
Through microscopic simulations and analytical derivations, we study different regimes for the semiconductor core, ranging from the hollow-core approximation, to the tubular-core nanowire appropriate for a semiconductor tube with an insulating core, to the solid-core nanowire with the characteristic dome-shaped radial profile for the electrostatic potential inside the semiconductor.
We compute the phase diagrams for the different models in cylindrical nanowires and find that MZMs typically coexist with CdGM analogs at zero energy, rendering them gapless. However, we also find topologically protected parameter regions or \textit{islands} with gapped MZMs.
In this sense, the most promising candidate to obtain topologically protected MZMs in a full-shell geometry is the nanowire with a tubular-shaped core.
Moving beyond pristine nanowires, we study the effect of mode mixing perturbations. On the one hand, mode mixing can gap CdGM analogs and open minigaps around existing MZMs. On the other hand and rather strikingly, mode mixing can act like a topological $p$-wave pairing between particle-hole Bogoliubov partners, and is therefore able to create new topologically protected MZMs in regions of the phase diagram that were originally trivial. As a result, the phase diagram is utterly transformed and exhibits protected MZMs in around half of the parameter space.

\end{abstract}

\maketitle

\section{Introduction}
\label{Sec:intro}

\begin{figure}
   \centering
   \includegraphics[width=\columnwidth]{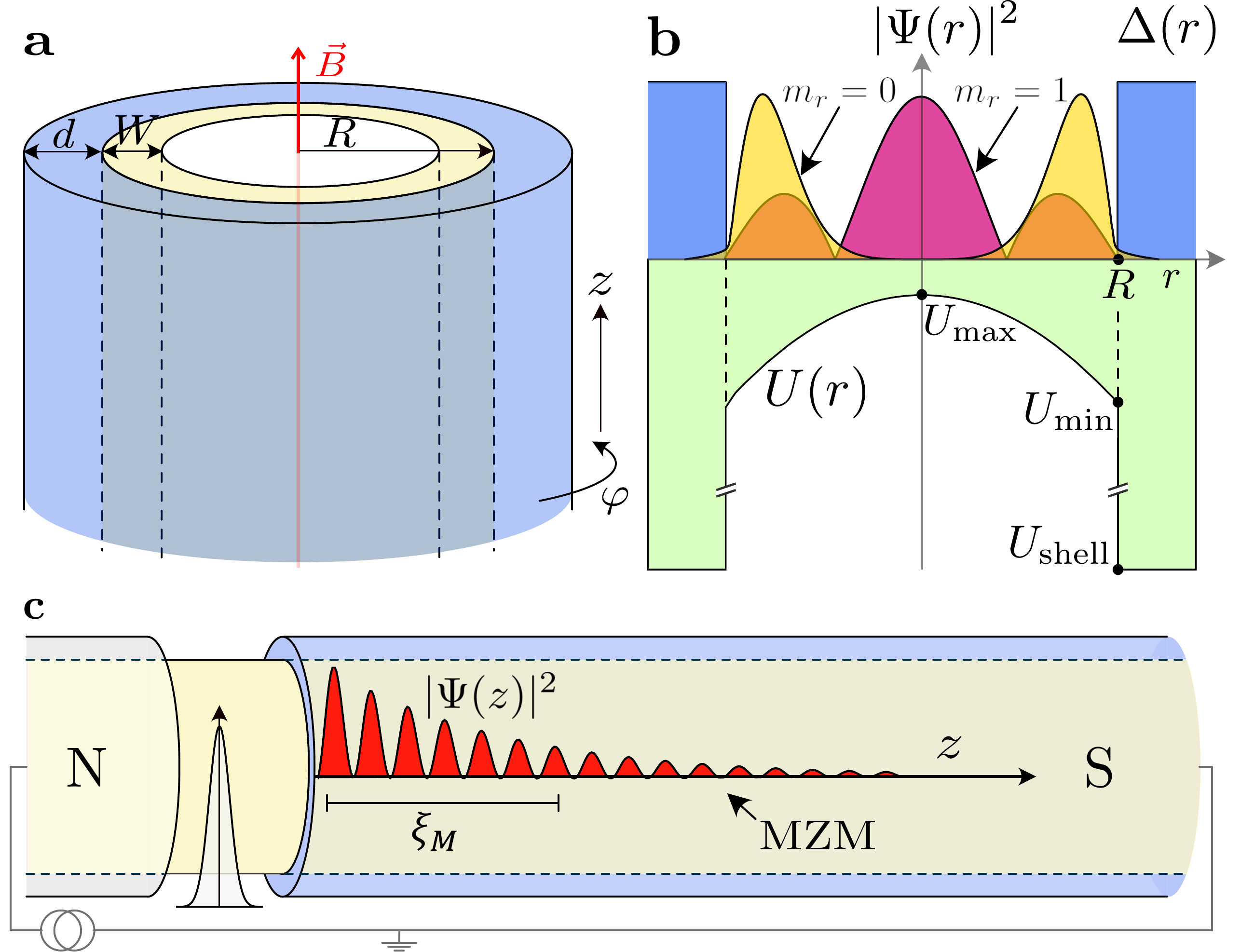}
   \caption{\textbf{Full-shell hybrid nanwowire.} (a) Sketch of a full-shell nanowire in a cylindrical approximation. A semiconductor core (yellow) of radius $R$ is surrounded by a thin superconductor shell (blue) of thickness $d$. For modellistic purposes, the core can have a tubular shape of thickness $W$. In an applied axial magnetic field $B$ the hybrid wire is threaded by a non-quantized magnetic flux $\Phi=\pi(R+d/2)^2B$. (b) Sketch of the radial dependence of the pairing amplitude $\Delta(r)$ (blue), electrostatic potential energy $U(r)$ (green) and typical subgap wavefunction density $|\Psi(r)|^2$ (yellow for the lowest radial mode $m_r=0$ and pink for the first radial mode $m_r=1$). The conduction-band bottom inside the semiconductor exhibits a dome-like radial profile with maximum value at the center, $U_{\rm{max}}$, and minimum value at the superconductor-semiconductor interface, $U_{\rm{min}}$. The electrostatic potential of the metallic shell is $|U_{\rm{shell}}|\gg |U_{\rm{min}}|$.  (c) Schematics of a full-shell nanowire-based normal-superconductor junction, characterized by a tunnel potential-barrier in the uncovered semiconductor region between the normal metal (N) and the full-shell wire (S). In red, Majorana bound state wavefunction at the end of the semi-infinite full-shell hybrid nanowire along the longitudinal direction. $\xi_{\rm M}$ is the Majorana localization length.}
   \label{fig:sketch}
\end{figure}

During the last decade there has been an intensive search for Majorana zero modes (MZMs) in the condensed matter community.
Arguably, the most explored platform for the realization and manipulation of MZMs is based on hybrid heterostructures combining superconductor and semiconductor materials~\cite{Lutchyn:Nat18}.
Although initial models for one-dimensional topological superconductivity in these systems were rather simple~\cite{Lutchyn:PRL10, Oreg:PRL10}, a fact that contributed to the strong interest and advances in the field, throughout the years it has been discovered that the experimental reality is far more complex than anticipated~\cite{Prada:NRP20}, substantially increasing the complexity of the required theoretical modelling, and making the creation and demonstration of MZMs subtle and subject to debate. % Nevertheless, the knowledge acquired by the community, from both the public and private sectors, about these systems in a relatively short period of time is impressive.

Among the current efforts to create and unequivocally demonstrate MZMs in hybrid heterostructures there exist several lines of research; we mention some of them. One is based on electrostatically defined quasi-one dimensional wires embedded in two-dimensional (2D) hybrid heterostructures~\cite{Kjaergaard:NatCom16, Suominen:PRL17, Lee:NanoLet19, Aghaee:PRB23, Ge:NatCom23}. These devices have the advantage that 2D electron gases in the semiconductor heterostructure present very high-mobilities and the electrostatic confinement reduces boundary effects as compared to traditional faceted nanowires, which should contribute to the reduction of the detrimental effects of disorder on MZMs. There have been claims~\cite{Aghaee:PRB23} of Majorana detection passing the topological gap protocol~\cite{Pikulin:A21} with topological minigaps up to $E_{\rm g}\sim 20-60$~$\mu$eV in Al/InAs devices, a prerequisite for experiments involving fusion and braiding of MZMs~\cite{RevModPhys.80.1083,Sarma:NPJ15,10.21468/SciPostPhys.3.3.021,10.1063/PT.3.4499,10.1063/PT.3.4499,BeenakkerReview_20,Zhou:NC22}. Planar Ge hole gases have also been considered recently~\cite{Laubscher:A23}. The generation of Majoranas in planar Josephson junctions defined on these heterostructures are another option~\cite{Banerjee:PRB23}, with the advantage of the extra control parameter of phase difference across the junction. Related to this last possibility, it has been proposed to create 1D topological superconductivity in double Josephson junctions in series. This proposal is based entirely on phase control and avoids the use of detrimental strong magnetic fields to drive the system into the topological transition~\cite{Lesser:PRB21, Lesser:PRB22, Lesser:JPD22, Luethi:A23}. A different, bottom-up approach consists of creating Majorana wires by concatenating hybrid quantum dots~\cite{Leijnse_PRB2012,Sau:NatCom12}, with the aim of directly implementing a Kitaev-chain model~\cite{Kitaev:PhysU01}. Recently, there have been important experimental progress with the observation of so-called poor man’s Majorana states in minimal QD chains~\cite{Dvir:N23, Sebastiaan:A23}.

Finally, the original (and probably most explored) line of reasearch around various types of hybrid nanowires has seen remarkable advances over the last years, from improved material and device aspects, to the proposal of alternative designs with advantageous properties over conventional nanowires. One example designed to avoid the application of external magnetic fields is the use of tri-partite nanowires, where a ferromagnetic insulator layer is added to the superconductor-semiconductor combo. This platform has been explored both experimentally~\cite{Vaitiekenas:NP21} and theoretically~\cite{Escribano:PRB21, Liu:PRB21, Woods:PRB21, Maiani:PRB21, Langbehn:PRB21, Escribano:npjQM22}. Here, we are interested on a different design variation, also famous for operating at small applied magnetic fields, known as full-shell hybrid nanowires.

Full-shell hybrid nanowires are semiconductor nanowires with strong spin-orbit (SO) coupling fully surrounded by a thin superconductor layer (or shell), see Fig.~\ref{fig:sketch}. They differ from conventional or partial-shell geometries, where the superconducting coating is limited to some facets of the nanowire~\cite{Mourik:S12,Krogstrup:NM15,Antipov:PRX18,Woods:PRB18,Winkler:PRB19}. The topological phase transition is triggered by an external magnetic flux threading the nanowire, whereas in the partial-shell devices following the original proposal \cite{Oreg:PRL10, Lutchyn:PRL10}, the trigger is the Zeeman effect. Full-shell hybrid nanowires came to the spotlight in 2020 thanks to an experimental-theoretical collaboration where signatures compatible with MZMs were observed~\cite{Vaitiekenas:S20}. Even though this work~\cite{Vaitiekenas:S20} quickly attracted the interest of the community~\cite{Woods:PRB19,Penaranda:PRR20,Vaitiekenas:PRB20,Kopasov:PSS20,Sabonis:PRL20, Paya:PRB23,Razmadze:A23,Giavaras:A23,Klausen:Nano23,Chen:A23}, the system's rich phenomenology remains largely unexplored.

One of the most striking phenomena in these wires is the so-called Little-Parks (LP) effect~\cite{Little:PRL62, Parks:PR64}. In the LP effect, the magnetic flux $\Phi$ through the section of the nanowire causes the superconducting phase in the shell to acquire a quantized winding around the nanowire axis. Winding number jumps are accompanied by a repeated suppression and recovery of the superconducting gap, forming so-called LP \textit{lobes} as a function of flux. These lobes are characterized by an integer number $n$ of fluxoids through the section. This effect has been measured in Al/InAs full-shell  nanowires~\cite{Vaitiekenas:PRB20, Razmadze:A23}, as well as in full-shell double nanowires~\cite{Vekris:SP21}.

Another important property of full-shell hybrid nanowires is the presence of a special type of subgap states inside the lobes. They are the result of the boundary condition imposed by the superconductor shell surrounding the semiconductor core, which induces a combination of normal and Andreev reflection at the core-shell interface, see Fig.~\ref{fig:sketch}(b). These subgap states can be regarded as the hybrid-nanowire \textit{analogs} of the Caroli--de Gennes--Matricon (CdGM) states in Abrikosov vortex lines of type-II superconductors~\cite{Caroli:PL64,Brun-Hansen:PLA68,Bardeen:PR69,Tinkham:04}. There exist, however, several differences between these CdGM analogs~\cite{Paya:PRB23} and their the Abrikosov counterparts. Their phenomenology is very rich and has been recently characterized in some detail~\cite{Kopasov:PRB20,Kopasov:PSS20,Paya:PRB23}.

The presence of an extra subgap state at zero energy of topological origin under certain circumstances and the corresponding topological phase diagram were discussed in Refs.~\onlinecite{Vaitiekenas:S20, Penaranda:PRR20}. In the search for MZMs, the full-shell design offers several advantages~\cite{Vaitiekenas:S20} as compared to partial-shell nanowires. For instance, the core of the wire is shielded from unwanted effects of the environment and surface disorder. They require smaller applied magnetic fields for the topological transition to happen, which is good to preserve the superconducting state of the parent superconductor shell. Moreover, MZMs are predicted to appear at very specific regions of parameter space, particularly at the \textit{odd} LP lobes. This might be useful to distinguish them from other unwanted trivial states.

However, full-shell nanowires also present some drawbacks~\cite{Vaitiekenas:S20,Woods:PRB19}, such as the fact that, once the hybrid wires have been grown, the electron density (and/or chemical potential) and the SO coupling are not tunable through direct gating, due to the metallic covering of the semiconductor core. These are essential parameters that determine the topological phase of the wire.
Moreover, using microscopic simulations, Ref.~\onlinecite{Woods:PRB19} predicted a small Rashba SO coupling in realistic Al/InAs full-shell wires, which lead to small and sparse chemical potential windows with non trivial topology, that could only be reached by tuning the radius of the wire. Other potential drawbacks of full-shell hybrid nanowires are shared with partial-shell ones, such as the detrimental effects of disorder, or the possible formation of quasi-Majoranas in the presence of smooth confinement at the wire ends or smooth potentials along the wire, a complication that still needs to be explored further in this system.

A couple of relevant works along this line experimentally demonstrated features compatible with the Majorana phenomenology but for devices in the trivial regime. In Ref.~\onlinecite{Valentini:S21}, a quantum dot formed at the junction between a local probe and the full-shell nanowire gives rise to robust zero-bias peaks in tunneling spectroscopy whose origin can be understood as non-topological subgap states in the Yu-Shiba-Rusinov regime. A subsequent theoretical explanation for the appearance of flattened parity crossings that simulate MZMs in this system was given in Ref.~\onlinecite{Escribano:PRB22}. Majorana-like Coulomb spectroscopy results were reported in Ref.~\onlinecite{Valentini:N22} without any concomitant zero bias peaks in tunnel spectroscopy. These observations were explained in terms of low-energy, longitudinally confined island states rather than overlapping MZMs.

In this work, we perform a comprehensive theoretical analysis of the phenomenology of long\footnote{Longer than the Majorana localization length, so that Majorana bound states at opposite ends do not hybridize}, full-shell hybrid nanowires in the topological phase, i.e., assuming that the wire parameters can fall within the topological regions of the phase diagram. We consider both pristine nanowires, modelled with a cylindrical approximation, and nanowires exhibiting mode-mixing perturbations that could be created by cross-section deformations or disorder. We otherwise ignore finite-length effects, disorder along the wire, or other sources of imperfections in the device. Our motivation is to clarify what behavior should be expected of MZMs in ideal conditions and what controls their degree of protection. We believe that this phenomenology can shed light on what is possible or realistic in present and future experiments.

We systematically explore different regimes of these hybrids, from an extreme situation where the wavefunction is localized at the core-shell interface, known as the hollow-core approximation~\cite{Vaitiekenas:S20}, to an intermediate situation where the wavefunction extends across a finite distance from the interface, which we dub the tubular-core model~\cite{Paya:PRB23}, all the way to the solid-core scenario, where the wavefunction can either extend homogeneously throughout the cross section or, more realistically, follow the typical non-homogeneous electrostatic potential $U(r)$ inside the core, see Fig.~\ref{fig:sketch}(b). This potential is a consequence of the band-bending imposed by the epitaxial core-shell Ohmic contact~\cite{Mikkelsen:PRX18,Chen:A23}. Through this step-by-step analysis we are able to explain the underlying reasons for the characteristics of the MZMs, while recovering and substantially extending previously known results~\cite{Vaitiekenas:S20,Penaranda:PRR20}.

We focus particularly on the signals produced by the presence of a Majorana bound state at the end of a semi-infinite full-shell hybrid nanowire as measured by local density of states (LDOS) or tunneling spectroscopy through a normal-superconductor junction, see Fig.~\ref{fig:sketch}(c). As a function of the threading flux, these quantities display a number of LP lobes and a subgap contribution coming from CdGM analogs. In the topological phase, zero energy peaks (ZEPs) of Majorana origin appear. We provide general analytical derivations and microscopic numerical simulations specifically for Al/InAs hybrids.

In the first part of the paper, and following Refs.~\onlinecite{Vaitiekenas:S20, Paya:PRB23}, we consider pristine full-shell nanowires modelled with a cylindrical approximation. Our main findings of this part are: (i) MZMs appear at odd LP lobes and typically coexist with CdGM analogs at zero energy.
(ii) We compute the topological phase diagrams for the different nanowire models. In general, we find topological regions with unprotected (gapless) MZMs, but also smaller parameter \textit{islands} with topologically protected MZMs (i.e., with a topological minigap). These islands happen only for low chemical potential. (iii) The flux interval with Majorana ZEPs in LDOS or $dI/dV$ at the $n=1$ LP lobe contains direct information on the spatial distribution of the Majorana wavefunction across the wire section. (iv) Tubular-core nanowires are specially suitable to create MZMs that can be spectrally separated from CdGM analogs.
(v) The subgap phenomenology of solid-core nanowires is rather complex due to the proliferation of CdGM subgap states, and crucially depends on whether one or more radial momentum subbands are occupied. With more than one, there is typically no topological minigap in this case.

%\editE{While point (i) and the topological phase diagram of the hollow-core approximation were already discussed in Refs.~\onlinecite{Vaitiekenas:S20, Penaranda:PRR20}, the rest of the points are original to this work.}

In the second part of the paper we introduce mode-mixing perturbations. In this case, our main findings are: (vi) Mode mixing acts like a topological $p$-wave pairing between particle-hole Bogoliubov partners, i.e., the time-reversed CdGM analogs crossing at zero energy. As a result, mode mixing is revealed to be a new mechanism for the formation of MZMs. (vii) The phenomenology of hexagonal cross-section full-shell nanowires is in general very similar to that of the cylindrical approximation. For some parameters, nevertheless, additional MZMs can arise with small minigaps in parameter regions where the cylindrical nanowire is trivial. (viii) In the presence of more generic disorder-induced mode-mixing perturbations, topological minigaps can open around previously gapless MZMs in odd lobes, and protected new MZMs can appear in even lobes. As a result, the phase diagram is utterly transformed by mode mixing, and exhibits protected MZMs in around half of the parameter space.

%\editE{The presence of mode-mixing perturbations and its effect on the topological phase diagram were previously analyzed in Refs.~\onlinecite{Vaitiekenas:S20,Penaranda:PRR20} for the solid-core case. In Ref. ~\onlinecite{Vaitiekenas:S20} a fully numerical approach for an hexagonal cross-section nanowire was employed, whereas here we introduce an analytical model that permits us to understand the underlying mechanisms of MZM formation, gap-openings and the shape of the topological regions. We moreover extend it to the tubular-core nanowire model.}

This paper is organized as follows. In Sec.~\ref{Sec:Cyl} we consider pristine full-shell hybrid nanowires, modelled with a cylindrical approximation. In Sec.~\ref{Sec:HCNw} we analyze the hollow-core approximation.
In Sec.~\ref{Sec:TCNw} we consider the tubular-core model for different semiconductor tube thicknesses. A useful approximation for the practical calculation of the tubular-core properties at a reduced computational cost is given in Sec.~\ref{Sec:MHCNw}.
In Sec.~\ref{Sec:SCNw} we consider the solid-core nanowire, taking into account a realistic band-bending profile at the core-shell interface. In Sec.~\ref{Sec:modemixing} we consider the effect of mode mixing perturbations and we conclude in Sec.~\ref{Sec:conclusions}. In App.~\ref{Ap:Model} we summarize the model we employ to characterize all the phenomenology of hybrid full-shell nanowires (including subsections of the LP effect of a diffusive shell [\ref{Ap:LPeffect}], the Bogoliubov-de Gennes Hamiltonian [\ref{Ap:Hamiltonian}], the quantum numbers [\ref{Ap:quantumnumbers}], the Green function numerical methods [\ref{Ap:Green}], the calculation of observables such as density of states, differential conductance and Majorana localization length [\ref{Ap:Obs}], and the modelling of mode-mixing perturbations [\ref{Ap:modemixing}]).
In App.~\ref{Ap:HCA_Topology} we provide additional details on the topological characterization of the hollow-core model.
The results presented in the main text correspond to the non-destructive LP regime. Results in the destructive regime are shown in App.~\ref{Ap:destructive}.

\section{Cylindrical approximation}
\label{Sec:Cyl}

\begin{figure*}
   \centering
   \includegraphics[width=\textwidth]{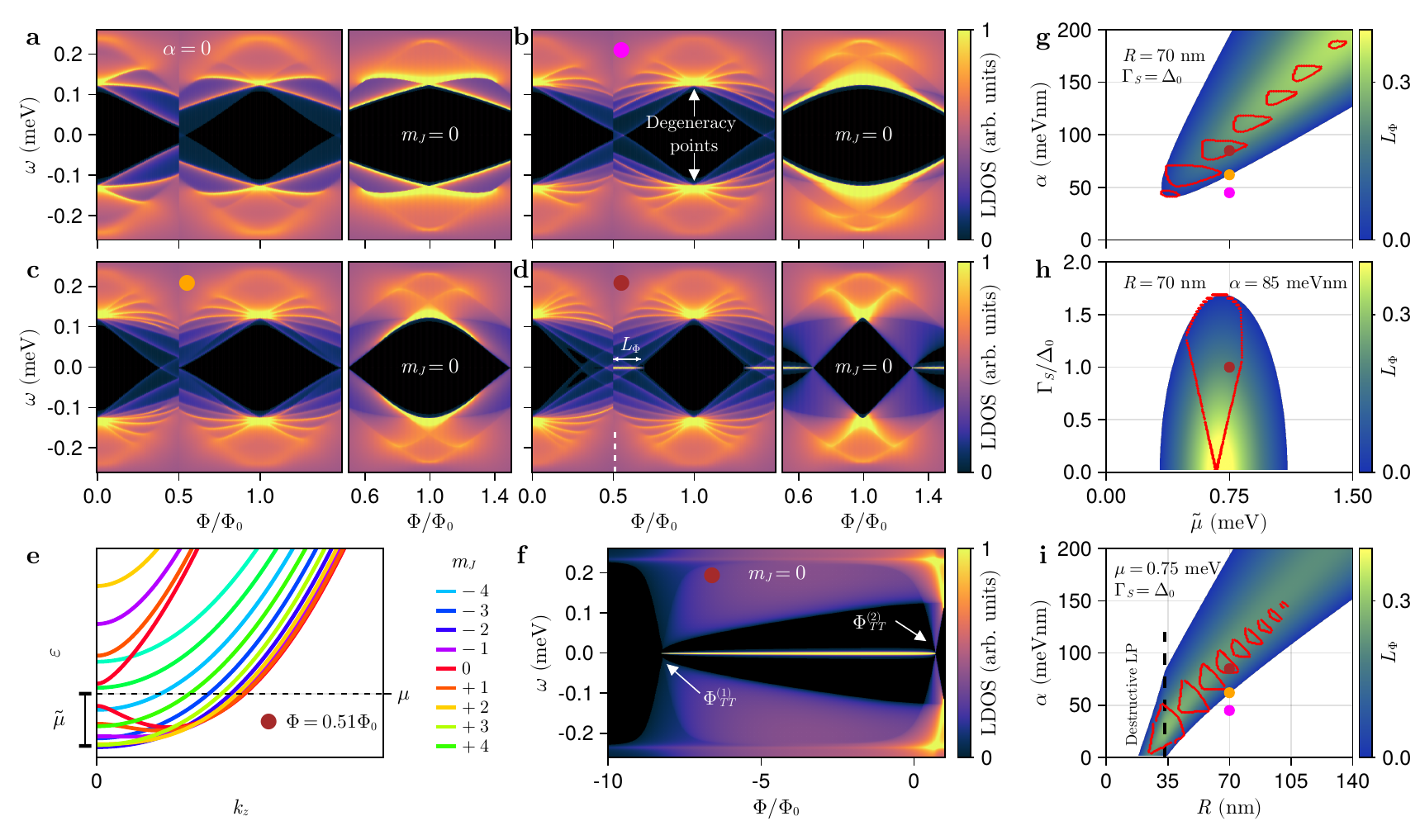}
   \caption{\textbf{Hollow-core model.} (a) Left panel: Local density of states (LDOS) at the end of a semi-infinite $R=70$~nm hollow-core nanowire (in arbitrary units) as a function of energy $\omega$ and applied normalized flux $\Phi/\Phi_0$. The right half of the $n=0$ and the full $n=1$ LP lobes are displayed. The wire is in the non-destructive LP regime. Right panel: LDOS of the $n=1$ lobe showing only the contribution of the $m_J=0$ subbands. Here, the SO coupling $\alpha=0$ and the Fermi energy $\tilde{\mu}=0.75$~meV. (b-d) Same as (a) but for different values of $\alpha$ marked by colored dots in (g). All the CdGM analogs coalesce at degeneracy points, located at the center of the LP lobes in the hollow-core approximation. (e) Band structure in the normal state (with decay rate into the shell $\Gamma_{\rm S}\rightarrow 0$) as a function of the longitudinal wave vector $k_z$, corresponding to the case (d) at $\Phi=0.51\Phi_0$. The number of occupied subbands depends on the Fermi energy $\tilde\mu$. The chemical potential $\mu$ is marked with a black dashed line. Different colors correspond to different values of the generalized angular momentum $m_J$ in the $n=1$ LP lobe. The $m_J=0$ sector gives rise to MZMs when proximitized. (f) $m_J=0$ LDOS  corresponding to the case (d) but fixing the fluxoid number to $n=1$ for all the magnetic-flux range displayed. The fluxes at which the $n=1$ metastable wire enters and exits the topological phase, $\Phi^{(i)}_{\rm TT}/\Phi_0$ with $i=1,2$ [see Eq.~\eqref{PhiTT}], are marked with arrows. (g) Topological phase diagram as a function of $\alpha$ and $\tilde\mu$ (for $R=70$~nm). The color scale represents the flux interval $L_{\Phi}$ of the left Majorana ZEP within the first LP lobe, see Eq.~\eqref{Lphi}. Red curves enclose regions of parameters, or \textit{islands}, with topologically protected Majorana ZEPs (i.e.,  MZMs with a minigap for some flux in the $n=1$ lobe). (h) Same as (g) but as a function of $\Gamma_{\rm S}$ and $\tilde\mu$ (for $\alpha=85$~meV\,nm). (i) Same as (g) but as a function of $\alpha$ and $R$ (with $\tilde\mu=0.75$~meV). For $R$ smaller than the vertical dashed line, the wire is in the destructive LP regime. Other parameters: $d=0$, $\Delta_0 = 0.23$~meV, $\xi_{\rm d} = 70$~nm, $m^*=0.023 m_e$, $\Gamma_{\rm S} = \Delta_0$, $g=0$ and $a_0=5$~nm.}
   \label{fig:HCA}
\end{figure*}

In this first part, we consider in general a full-shell hybrid nanowire like the one depicted in Fig.~\ref{fig:sketch}(a), with a semiconductor core of radius $R$ and a thin superconductor shell of thickness $d$. We assume for simplicity that the hybrid wire has cylindrical symmetry, with radial coordinate $r$, azimuthal angle $\varphi$ and axial coordinate along the $z$ direction. In Sec.~\ref{Sec:modemixing} we will show that this is a very good approximation for a more realistic hexagonal-shaped nanowire, and we will discuss the effect of more dramatic cross-section deformations. The full-shell nanowire is threaded by a magnetic field $\vec{B}=B\hat{z}$ that gives rise to a flux
\beqa
\label{flux}
\Phi &=& \pi R_\mathrm{LP}^2 B,\\
R_\mathrm{LP} &=&R + d/2.\nonumber
\eeqa
Note that $\Phi$ is taken at the mean radius $R_\mathrm{LP}$ of the shell. The methodology to analyze this system can be found in App.~\ref{Ap:Model}. In particular, the effective BdG Hamiltonian is given in Eq.~\eqref{solidrot}. It is expressed in terms of the flux and geometrical parameters, the decay rate from the core into the superconductor $\Gamma_{\rm S}$\footnote{See discussion in App. \ref{Ap:Hamiltonian} for our definition of a frequency-dependent BdG effective Hamiltonian and its dependence on the decay rate $\Gamma_{\rm S}$.}, the intrinsic parameters such as the effective mass $m^*$, the chemical potential $\mu$, the SO coupling $\alpha$ or the g factor, and on the generalized angular momentum quantum number $m_J$, see App.~\ref{Ap:quantumnumbers}. This quantum number labels the different transverse subbands of the wire and takes half-integer or integer values for even and odd LP lobes, respectively. Even (odd) lobes are characterized by an even (odd) integer number $n$ of superconductor phase windings or, equivalently, fluxoid number.

In the following subsections we consider different models for the core, from the somewhat artificial hollow-core approximation to the solid-core case.

\subsection{Hollow-core model}
\label{Sec:HCNw}

In this first section, we examine the simplest approximation to the above full-shell hybrid nanowire model, by assuming that all the semiconductor charge density is located at the interface with the superconductor shell \cite{Vaitiekenas:S20}. This is called the hollow-core approximation, and corresponds to fixing $r=R$, $d=0$ in Eqs.~\eqref{solidrot} and \eqref{shelfenergy}. Moreover, we define $\tilde\mu\equiv\mu-U(R)-\langle p_r^2\rangle/2m^*$, where $\langle p_r^2\rangle/2m^*$ represents the radial confinement energy and, thus, $\tilde\mu$ represents the Fermi energy (or energy difference between the highest and lowest occupied single-particle states at zero temperature). Even though the hollow-core model is a drastic approximation, we still take into account (i) the effect of the magnetic flux on the superconducting shell (the LP effect), (ii) the proximity effect on the core subbands with well-defined angular momentum $m_J$, and (iii) the effect of the magnetic flux on the core subbands. As we shall see later on, the hollow-core model is a very coarse approximation when compared to the results obtained in more realistic full-shell hybrid nanowires. However, it is very useful to understand the basic phenomenology of these wires and the results of more sophisticated models.

In the left panels of Fig.~\ref{fig:HCA}(a-d) we plot the LDOS at the end of a semi-infinite hollow-core nanowire of radius $R=70$~nm as a function of energy $\omega$ and normalized applied flux $\Phi/\Phi_0$, where $\Phi_0$ is the superconducting flux quantum. The LDOS is $\pm \omega$ symmetric as corresponds to a BdG Hamiltonian. It is also $\pm \Phi$ symmetric (not shown). Observe the LP modulation with flux of the parent-gap edge (i.e., the gap of the shell) that defines the different lobes. Here we show only half of the $n=0$ LP lobe and the complete $n=1$ lobe. For this $R$, the shell is in the non-destructive LP regime, i.e., the parent gap  does not close between lobes. In the hollow-core approximation all even and all odd lobes display the same LDOS, respectively, and the LDOS within each lobe is symmetric respect to its center. The LP lobe outline is smooth in $\omega$, which is a consequence of using a diffusive self energy for the shell.

In Fig.~\ref{fig:HCA}(a) we consider the case with $\alpha=0$, i.e., in the absence of SO coupling. We see a number of bright features below the parent gap in the different lobes. These are the so-called CdGM analog states analyzed in detail Ref.~\onlinecite{Paya:PRB23}. They are Van Hove singularities, two for each $m_J$, that are induced by the superconductor shell on the propagating core states. For the Fermi energy $\tilde\mu=0.75$~meV chosen here, there are four populated angular momentum subbands in the even lobes, with quantum numbers $m_J=\pm1/2, \pm 3/2$. In the odd lobes, the populated subbands are five, $m_J=0, \pm 1, \pm 2$. CdGM analogs corresponding to $+m_J$ ($-m_J$) disperse with flux with positive (negative) slope. Close to the lobe centers (or in the limit of small coupling to the superconductor) they disperse linearly, but level repulsion makes them bend downwards when their energy approaches that of the parent gap at the lobe edges. Note that the CdGM analogs coalesce at the lobe centers, where the semiconductor charge density is threaded by an integer number of flux quanta $n\Phi_0$, into what we dub the \textit{degeneracy} points.
This happens because, for $\Phi=n\Phi_0$ and in the limit $d\rightarrow 0$, the terms that depend on $n$ and $\Phi$ in the Hamiltonian of Eq.~\eqref{solidrot} cancel, and thus the system is equivalent to that at $\Phi=0$. Moreover, at $\Phi=0$ all the CdGM analogs experience the same induced gap, since they have the same spatial density at the core-shell interface.

The distance between the Van Hove singularities and the parent-gap edge is controlled by the superconductor-semiconductor coupling $\Gamma_{\rm S}$, and defines the true induced gap of the hybrid wire (black color in the LDOS), which is typically smaller than the parent gap. In general, we give $\Gamma_{\rm S}$ in units of the zero-field gap, denoted by $\Delta_0$.
In the limit of large $\Gamma_{\rm S}$, the superconducting proximity effect is so strong that all the CdGM states are pushed towards the parent-gap edge, resulting in empty lobes in the LDOS. Conversely, in the limit $\Gamma_{\rm S}\rightarrow 0$, the lobes are filled with CdGMs analogs. For an intermediate value, such in the case of Fig.~\ref{fig:HCA}(a-d), there tends to be a true induced gap below the degeneracy points (with a diamond-like shape) and a gapless region around the lobe edges (at half-integer normalized flux).

The effect of a finite SO interaction is analyzed in Figs.~\ref{fig:HCA}(b-d) for the three increasing values of SO coupling marked by colored dots in Fig.~\ref{fig:HCA}(g). The Fermi energy is the same as in  Fig.~\ref{fig:HCA}(a). A growing $\alpha$ has two distinct effects. On the one hand the nanowire experiences an extra doping that increases the number of occupied subbands and thus of CdGM analog features in LDOS. This effect stems from term of Eq.~\eqref{LutchynC} in App.~\ref{Ap:HCA_Topology}. On the other hand, the Van Hove singularities away from the degeneracy points split in energy for different spin quantum number, whereas in Fig.~\ref{fig:HCA}(a) the $\alpha=0$ CdGM analogs are spin-degenerate.

The SO-induced splitting in the $m_J=0$ sector is responsible for the emergence of topological MZMs. To see this, in the right panels of Fig.~\ref{fig:HCA}(a-d) we show the contribution to the first-lobe LDOS coming only from the $m_J=0$ subbands. For $\alpha=0$, we see that there are two Van Hove singularities that cross at the center of the lobe for $\omega>0$. As we turn on $\alpha$, right panel of Fig.~\ref{fig:HCA}(b), these features transform into one bright dome-shaped CdGM state at lower energy, and some additional, more-complicated LDOS structure above. As we increase $\alpha$ further, right panel of Fig.~\ref{fig:HCA}(c), the dome-shaped CdGM state crosses zero energy at the lobe edges. This corresponds to a topological phase transition, associated with a zero-energy crossing at $k_z=0$ of the $m_J=0$ subband in the band structure of the corresponding bulk nanowire. Increasing $\alpha$ even further, right panel of Fig.~\ref{fig:HCA}(d), a ZEP appears across a finite flux interval at each edge of the lobe.
The Majorana bound state responsible for these ZEPs is localized at the edge of the semi-infinity nanowire, in red in Fig.~\ref{fig:sketch}(c). In the $m_J=0$ LDOS of the right panel of Fig.~\ref{fig:HCA}(d), it is possible to see the gap closing and reopening at the particular flux where the topological transition takes place. There is a topological minigap in the $m_J=0$ sector (a finite distance between the MZM and the $m_J=0$ induced gap). However, when the total LDOS is considered with all the populated $m_J$, see left panel of Fig.~\ref{fig:HCA}(d), there is no true minigap throughout most of the ZEP, except at the very tip where it emerges in this particular case.
Technically, then, only this tip is topologically protected. In the following, we will dub nanowires with a Majorana ZEP as \emph{topological}, regardless of whether they have a topological minigap or not. In addition, nanowires with a ZEP that does not coexist at zero energy with any gapless CdGM analogs will be dubbed \emph{topologically protected}. Topologically protected wires are only possible if, as $\Phi$ approaches the lobe edges, the $m_J=0$ CdGM analog reaches zero energy before any other populated $m_J$. This is a more restrictive condition than simply having a MZM somewhere in the lobe.

Even though we have just seen that the effect of the SO coupling is dramatic in the $m_J=0$ sector (changing the shape of the Van Hove singularities, driving the topological phase transition and giving rise to the Majorana ZEPs), we note that $\alpha$ has a small effect on the rest of the $m_J\neq 0$ CdGM analogs (other than increasing their number due to the $\alpha$-mediated self doping). This was also noted in Ref.~\onlinecite{Paya:PRB23}.

The band structure for an infinite wire corresponding to Fig.~\ref{fig:HCA}(d) is shown in Fig.~\ref{fig:HCA}(e). It is calculated for simplicity in the normal state, $\Gamma_{\rm S}\rightarrow 0$, and only electron bands are shown. It is evaluated at $\Phi=0.51\Phi_0$, i.e., at the left edge of the $n=1$ LP lobe. The colors designate pairs of bands with the same $m_J$ quantum number. Each pair is separated by a finite energy gap\footnote{The two subbands of each pair have opposite spin quantum numbers $m_s=\pm 1/2$, and orbital angular momentum quantum numbers $m_l=m_J-1$ and $m_l=m_J$, see App.~\ref{Ap:quantumnumbers} for a definition of the quantum numbers.}. The subband pair that gives rise to MZMs, $m_J=0$, is colored in red. In the topological phase, the chemical potential $\mu$, marked by a dashed black line, has to be between the two subbands of the $m_J=0$ pair.

The appearance of the two ZEPs in Fig.~\ref{fig:HCA}(d) are preceded by a topological phase transition inside the $n=1$ LP lobe. However, they die out abruptly at the lobe edges where the fluxoid number changes from odd to even. The disappearance of the MZMs is not mediated by a band inversion, as in a usual topological phase transition, but rather by a first-order phase transition of the fluxoid number $n$. If one could fix the fluxoid to $n=1$ outside the first lobe (so that the system remains in a metastable state), we would be able to follow the MZMs and their eventual disappearance at another topological band inversion. We illustrate this for the left ZEP of Fig.~\ref{fig:HCA}(d). In Fig.~\ref{fig:HCA}(f) we plot the $m_J=0$ contribution to the metastable LDOS at a fixed $n=1$ for a wide range of flux beyond the first lobe. While the \textit{right} topological phase transition happens for $\Phi\approx 0.7\Phi_0$ (within the first lobe), there is a second topological band inversion at $\Phi\approx -8\Phi_0$.

It is possible to get an analytical expression for the flux values at which the topological transitions happen as a function of the wire parameters,
\begin{widetext}
\begin{equation}
    \frac{\Phi^{(i)}_{\rm TT}}{\Phi_0} = \left(1 \pm \sqrt{1 + 4 m^* R \left(\alpha + 2 m^* R\alpha^2 + 2 R \tilde\mu\right) \pm 4 R \sqrt{m^* \left[ \left(1 + 2 m^* R \alpha \right)^2 \left( m^* \alpha^2 + 2 \tilde\mu \right) - 4 m^* R^2 \Gamma_{\rm S}^2 \right]}}\right) \left(\frac{R_{\rm LP}}{R}\right)^2,
    \label{PhiTT}
\end{equation}
\end{widetext}
expressed in $\hbar=1$ units. This expression if valid for a finite $d$, contained in $R_{\rm LP}$ \footnote{ It is not possible to get an explicit analytical expression when the Zeeman term is included in the Hamiltonian.}. Note that to obtain this equation we take $\omega\rightarrow 0$ in the shell self energy, see Eq.~\eqref{Eq:SEomega0}. Equation~\eqref{PhiTT} has in general four solutions $i \in [1, 4]$ ($i = 1$ corresponds to $-+$, $i = 2$ to $- -$, $i = 3$ to $+ -$ and $i = 4$ to $+ +$), see App.~\ref{Ap:HCA_Topology} for a discussion. If $\tilde\mu<\tilde\mu_{\rm c}$, where
\begin{equation}
\label{muc}
\tilde\mu_{\rm c}=\frac{2m^*R^2\Gamma_{\rm S}^2}{(1+2m^*R\alpha)^2}-\frac{m^*\alpha^2}{2},
\end{equation}
the solutions are complex and the wire is in the trivial phase. If $\tilde\mu\ge\tilde\mu_{\rm c}$, the four solutions are real, a pair of them ($i=1,2$) corresponding to the left ZEP and the other two ($i=3,4$) to right ZEP. In Fig.~\ref{fig:HCA}(f), $\Phi^{(1)}_{\rm TT}$ and $\Phi^{(2)}_{\rm TT}$ are marked with arrows. Note that only if the inner solutions (those closer to $\Phi=\Phi_0$ with $i=2,3$) are between 0.5 and 1.5$\Phi_0$, i.e., within the first lobe, the hybrid wire will be in the topological phase for a particular set of parameters.
It is possible to define a MZM flux interval
\begin{equation}
\label{Lphi}
    L_\Phi \equiv \mathrm{clip}\left(\frac{\Phi^{(2)}_{\rm TT}}{\Phi_0} - \frac{1}{2}, [0,1]\right),
\end{equation}
where $\mathrm{clip}(x, [a,b]) = \min(\max(x, a), b)$. This is the extension in flux of the left Majorana ZEP within the $n=1$ LDOS, see Fig.~\ref{fig:HCA}(d). It has no units and, for $d=0$, it is bounded between 0 and 0.5. A topological phase is defined by $L_\Phi>0$.

The corresponding topological phase diagram of the $R=70$~nm hollow-core nanowire is shown in Fig.~\ref{fig:HCA}(g) as a function of constant $\alpha$ and $\tilde\mu$. Here $\Gamma_{\rm{S}}/\Delta_0=1$. The colored area shows the parameter region where odd lobes contain MZMs. Its boundary can be obtained analytically from $L_\Phi=0$ using Eqs. \eqref{Lphi} and \eqref{PhiTT}. The three SO couplings considered in Fig.~\ref{fig:HCA}(b-d) are marked with colored dots, for a trivial case (pink), at the topological phase transition (orange) and for a topological case (brown). The color scale represents the flux interval $L_{\Phi}$ of the left ZEP, which is larger at the center of the wedge-shaped topological region, and goes to zero the boundaries. An equivalent phase diagram, but for fixed $\alpha=85$~meV\,nm and as a function of $\Gamma_{\rm{S}}$ and $\tilde\mu$ is given in Fig.~\ref{fig:HCA}(h). If the decay rate to the superconductor $\Gamma_{\rm{S}}$ is too large, there cannot be MZMs because the proximity effect is so strong that all the subgap states are pushed to the parent-gap edge, including $m_J=0$, and the topological phase transitions cannot occur inside the odd lobes. Red curves inside the topological region in these diagrams enclose smaller \textit{topologically protected islands}, i.e., containing a  MZM with a topological minigap within odd lobes. We see several islands in Fig.~\ref{fig:HCA}(g), which correspond to an increasing number of populated $m_J$ subbands. As we enter a protected island from below, the $m_J=0$ CdGM analog overtakes the highest-occupied $m_J$ in its shift towards zero energy, so the tip of the ZEP does not coexist with any other CdGM at zero energy. Exiting an island from above, a new higher $m_J$ becomes occupied, introducing a new LDOS contribution that covers the tip of the ZEP.

For the $R=70$~nm hollow-core nanowire analyzed here, the values of the SO coupling needed to enter the topological phase are very strong ($\alpha\gtrsim \alpha_{\rm min} \approx 50$~meV\,nm). As we go away from the hollow-core approximation into the more realistic solid-core model, we will see that the minimum SO coupling value is reduced. Additionally, it is possible to get a topological phase for smaller $\alpha$ values by decreasing the nanowire radius $R$. In Fig.~\ref{fig:HCA}(i) we plot the phase diagram as a function of $\alpha$ and $R$. Observe that for $R\approx 30$~nm, which is still a realistic nanowire radius, the minimum value of the SO coupling to enter the topological phase vanishes, $\alpha_{\rm min}\rightarrow 0$. At these small values of $R$, the wire enters the non-destructive LP regime, see Ap. \ref{Ap:LPeffect}. In this case the flux interval $L_\Phi$ is shortened from the left, so that the $1/2$ in Eq. \eqref{Lphi} needs to be replaced by the leftmost flux of the first lobe.

\begin{figure*}
\centering
\includegraphics[width=\textwidth]{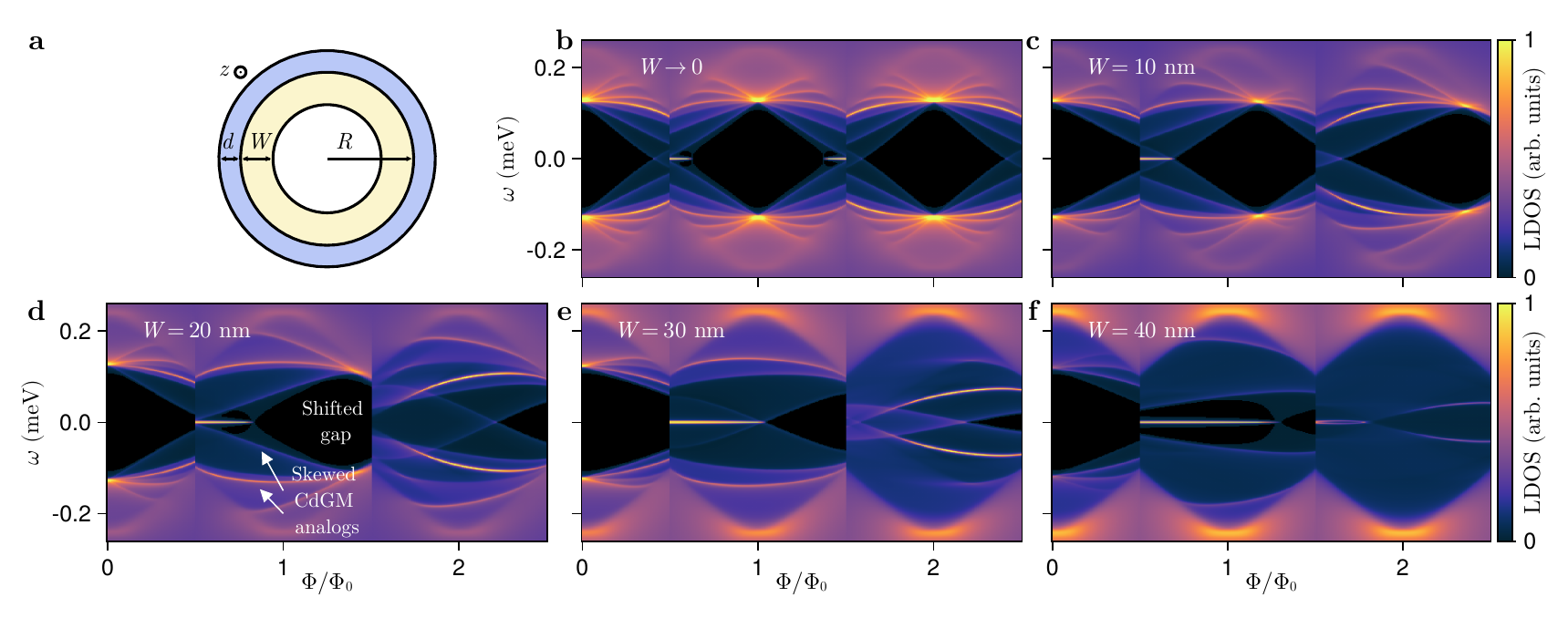}
\caption{\textbf{Tubular-core model.}
(a) Sketch of a tubular-core nanowire with shell thickness $d$, semiconductor-core radius $R$ and thickness $W$. (b-f) Local density of states (LDOS) at the end of a semi-infinite $R=70$~nm, $d\rightarrow 0$ tubular-core nanowire (in arbitrary units) as a function of energy $\omega$ and applied normalized flux $\Phi/\Phi_0$, displaying half of the $n=0$, and the full $n=1,2$ LP lobes. Panel (b) corresponds to the hollow-core approximation ($W\rightarrow 0$) and from (c) to (f) the thickness of the semiconductor tube increases in steps of $10$~nm. Parameters: $\alpha$ and $\mu$ are chosen to remain approximately at the same spot within the wedged-shaped topological region of the different-$W$ topological phase diagrams, see upper row of Fig. \ref{fig:TCMPD}. $\Gamma_{\rm S}$ is chosen so that the degeneracy points in the $n=0$ LP lobe remain fixed around $\omega = \pm 0.1$~meV, see lower row of Fig. \ref{fig:TCMPD}. For example, in panel (f) $\alpha=20$~meV\,nm, $\mu=7.1$~meV, $\Gamma_{\rm S}=48\Delta_0$, and $E_{\rm g}=52$~$\mu$eV at $\Phi=0.9\Phi_0$. Other parameters as in Fig.~\ref{fig:HCA}.}
\label{fig:TCMpos}
\end{figure*}

\begin{figure*}
\includegraphics[width=\textwidth]{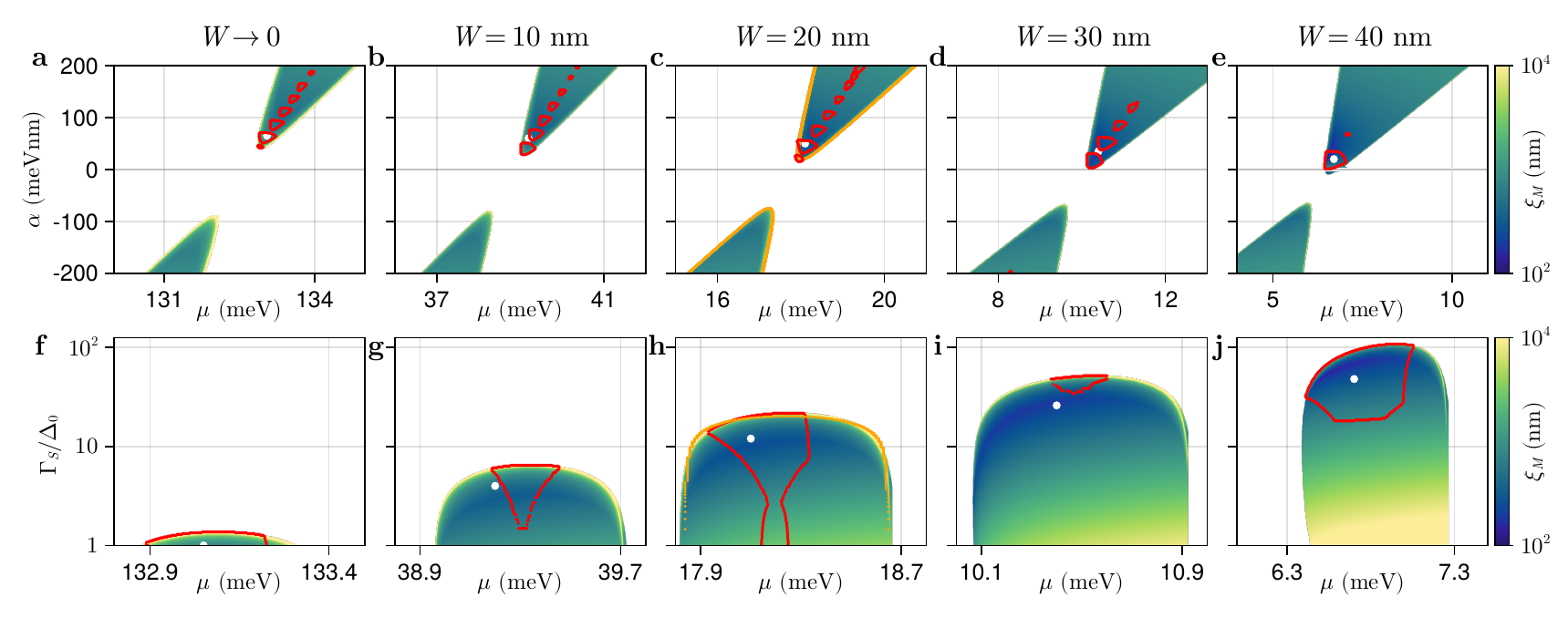}
\caption{ \textbf{Topological phase diagram for the tubular-core model}. (a-e) Topological phase diagrams as a function of SO coupling $\alpha$ and chemical potential $\mu$ for tubular-core nanowires with $R=70$~nm, $d\rightarrow 0$ and increasing values of tube thickness $W$. In each panel $\Gamma_{\rm S}$ corresponds to the white dot in the panel below. The color bar represents the Majorana localization length $\xi_{\rm M}$ at $\Phi=0.51\Phi_0$. The white dot in each panel corresponds to the parameters used in Fig.~\ref{fig:TCMpos}(b-f), respectively. The red islands contain parameters with topologically protected MZMs. (f-j) Same as (a-e) but as a function of decay rate $\Gamma_{\rm S}$ and chemical potential $\mu$, with $\alpha$ fixed to the white dot in the panel above. The solid orange curves in (c, h) (that contours the topological regions) have been calculated using the modified hollow-core approximation with $R_{\rm av}=59.5$~nm. For these curves, $\tilde\mu\in [-2.7,3.3]$~meV and $\Gamma^{\rm av}_{\rm S} / \Delta_0 \in [0,2.1]$. Rest of parameters as in Fig.~\ref{fig:HCA}.}
\label{fig:TCMPD}
\end{figure*}

\subsection{Tubular-core model}
\label{Sec:TCNw}

We now consider a full-shell hybrid nanowire where the semiconductor core is still hollow but has a finite thickness $W$, see yellow region in Fig.~\ref{fig:TCMpos}(a). This is what we call the tubular-core model, where for simplicity we take the SO coupling $\alpha$ to be spatially uniform in the core and $U(r)=0$ in the Hamiltonian~\eqref{solidrot} for the $r$-values in the yellow region. This approximation is justified if the semiconductor tube is not very thick. On the one hand, this model is useful to theoretically understand the fate of the CdGM analogs, as well as the MZMs, as we allow the charge density inside the core to spread away from the superconductor-semiconductor interface, thus generalizing the idealized hollow-core approximation analyzed in Sec.~\ref{Sec:HCNw}. On the other hand, this can also be a good model to describe a real full-shell tubular nanowire. This could be fabricated, for example, using a core-shell nanowire with an outer semiconductor shell and an inner insulating core, and then covered all around with a superconductor shell as before [blue in Fig.~\ref{fig:TCMpos}(a)].

In Fig.~\ref{fig:TCMpos}(b-f) we show the LDOS analogous to that in Fig.~\ref{fig:HCA}(d), i.e., in the topological phase. We now display also the $n=2$ lobe and gradually increase the semiconductor thickness in steps of 10~nm, from the hollow-core limit $W=0$ in Fig.~\ref{fig:TCMpos}(b) to $W=40$~nm in Fig.~\ref{fig:TCMpos}(f). We change the SO coupling and the chemical potential between panels as shown by the white dots in the upper row of Fig. \ref{fig:TCMPD}, following the downward movement of the topological region as  $W$ is increased. For example, $\alpha=70$~meV\,nm for $W\rightarrow 0$, but $\alpha=20$~meV\,nm for $W=40$~nm. As we increase $W$, smaller values of the SO coupling are required to enter the topological phase. Note that the parent-gap edge remains unchanged between panels, as its shape depends on the shell geometrical parameters that we choose as in Fig.~\ref{fig:HCA}. For a given core-shell coupling $\Gamma_{\rm S}$, the proximity effect depends on the core width, and is much stronger for a thin tubular semiconductor, and much weaker for a solid core one. This is in turn reflected on the energy position (and dispersion) of the CdGM analog states and the induced gap. For these simulations we take $\Gamma_{\rm S}$ in each panel so that the degeneracy points in the $n=0$ lobe are located approximately at $\omega=\pm 0.1$~meV, which implies an increasing coupling that goes from $\Gamma_{\rm S}= \Delta_0$ for $W\rightarrow 0$ to $\Gamma_{\rm S}=48 \Delta_0$ for $W=40$~nm, see white dots in the lower row of Fig. \ref{fig:TCMpos}.

For $W=0$, Fig.~\ref{fig:TCMpos}(b), the LDOS within each lobe is symmetric with respect to the lobe center, as corresponds to the hollow-core approximation. Two small Majorana ZEPs appear at the $n=1$ lobe edges. As we increase the core thickness $W$, the symmetry is lost and two remarkable things happen: (i) The degeneracy points shift to the right \footnote{Note that the degeneracy points shift to the left for negative lobes and fluxes, since the LDOS is symmetric with respect to $\Phi=0$, not shown here.}, to larger values of magnetic flux within each $n\neq 0$ lobe, dragging with them the corresponding CdGM analogs \cite{Paya:PRB23}. This is turn produces skewed-shaped Van Hove singularities and a shifted induced gap, see for instance Fig.~\ref{fig:TCMpos}(d). With the parameters of Fig.~\ref{fig:TCMpos}, the $n=1$ degeneracy points exit the first lobe and are not visible anymore for thicknesses $W\gtrsim 30$~nm, while the shifted induced gap disappears for $W>40$~nm. For the second lobe, the shift towards larger values of flux happens twice as quickly. (ii) The Majorana ZEPs also get shifted to larger values of flux. This implies that the right ZEP quickly disappears from the first lobe as we increase $W$, and the left one starts covering a wider range of magnetic flux until it eventually extends across the whole lobe. With the parameters of Fig.~\ref{fig:TCMpos}, this happens for $W> 40$~nm. Note that the MZMs in the tubular-core nanowire can display a sizable topological minigap for a large flux window.
For instance, for $W=20$~nm in Fig.~\ref{fig:TCMpos}(d), with $\alpha=50$~meV\,nm, there is minigap close to the rightmost tip of the ZEP with a maximum $E_{\rm g}=30$~$\mu$eV at $\Phi=0.72\Phi_0$. For $W=40$~nm in Fig.~\ref{fig:TCMpos}(f), with $\alpha=20$~meV\,nm, there is minigap all across the MZM flux interval with a maximum $E_{\rm g}=52$~$\mu$eV at $\Phi=0.9\Phi_0$. However, the $W=30$~nm panel happens to have no minigap because the parameters we have chosen in the topological phase diagram fall between two islands, see white dot in Fig. \ref{fig:TCMPD}(d).

It is possible to understand the shift with $W$ of all the subgap features towards larger values of flux in terms of the wavefunction radial distribution of the occupied modes inside the semiconducting core. This was already analyzed for the CdGM analogs in Ref.~\onlinecite{Paya:PRB23} for $\alpha=0$, but the same reasoning can be applied here to the Majorana ZEPs. The key argument is that, for a fixed $W$ (and not too large $\mu$), all the populated $m_J$ subbands are in the lowest radial subband (smallest radial momentum, with radial quantum number $m_r=0$), and have approximately the same radial profile: a standing wave that is zero at the inner and outer radii of the tube and that is maximum at an average radius $R_\mathrm{av} = \langle r\rangle$ \footnote{This is true as long as the degeneracy points are well defined, which for Fig.~\ref{fig:TCMpos} happens for all $W/R\lesssim 30-40\%$. As $W$ increases approaching $R$, different $m_J$ subbands start to have different $R_{\rm av}$ values, the degeneracy points exist the first lobe and they eventually dissolve. See Ref.~\onlinecite{Paya:PRB23} for a discussion of this effect.}, see Fig. 6(e) in Ref.~\onlinecite{Paya:PRB23} and Fig.~\ref{fig:MHC}(b) here. In the presence of a threading magnetic field, the flux experienced by the superconductor shell $\Phi=\pi R_{\rm LP}^2B$ is controlled by the LP radius $R_{\rm LP}$, which determines the period of the LP lobes. However, the effective flux experienced by the subgap features, $\pi R_{\rm av}^2B$, is instead controlled by $R_\mathrm{av}$, i.e, as if the spread-out CdGM wavefunctions were concentrated at $R_\mathrm{av}\leq R_{\rm LP}$. For $W=0$, $R_\mathrm{av}\approx R_{\rm LP}$ and thus the degeneracy points occur at the lobe centers. But as $W$ increases, $R_\mathrm{av}$ becomes smaller than $R_{\rm LP}$. Thus, the necessary flux for the CdGM wavefunctions to enclose an integer number of flux quanta, $\Phi_\mathrm{dp}=\pi R_\mathrm{LP}^2B_{\rm{dp}}$, increases, producing the shift. For a finite $W$, the degeneracy-point flux in the $n=1$ lobe is~\cite{Paya:PRB23}
\beq
\label{shift}
\frac{\Phi_\mathrm{dp}}{\Phi_0} = \left(\frac{R_\mathrm{LP}}{R_\mathrm{av}}\right)^2.
\eeq

In Fig.~\ref{fig:TCMPD} we show the topological phase diagrams for the tubular-core model of $R=70$~nm (and $d\rightarrow 0$) and different values of $W$. The upper row analyzes the behavior versus constant $\alpha$ and $\mu$, and the lower row the behavior versus $\Gamma_{\rm S}$ and $\mu$. As mentioned, the white dots in each panel correspond to the parameters used in each LDOS of Fig.~\ref{fig:TCMpos}. The color scale here represents the Majorana localization length, computed using the methodology of App.~\ref{Ap:Obs} and defined as the decay length of the Majorana bound state, see Fig. \ref{fig:sketch}(c). Note that in the upper row we now consider both positive and negative values of $\alpha$, see discussion of App.~\ref{Ap:HCA_Topology}. In principle, for a Rashba SO coupling produced by the Al/InAs band bending of Fig. \ref{fig:sketch}(c), $\alpha$ should be positive, since the electric field is radial and pointing inwards, towards the wire center, and the SO coupling is proportional to minus the electric field [see Eq. \eqref{Eq:SOC}]. However, we also include in our analysis the possibility of a negative $\alpha$ for completeness, since the presence of strain at the interface or a different source of SO coupling in other materials could lead to a different sign. The topological phase boundaries for negative and positive values of $\alpha$ (though not the Majorana localization lengths or minigaps) are symmetric respect to a point $\alpha_{\rm c} = -1 / (2m^*R_\mathrm{av}) < 0$, see Eq.~\eqref{alphac}. This implies that, in absolute value, it is necessary to have a larger SO coupling to enter the topological phase for negative $\alpha$ than for a positive one. Another interesting feature is that, as we increase $W$, the wedged-shaped topological regions move towards smaller values of $|\alpha|$, making it easier to enter the topological phase. Actually, for $W=40$~nm in Fig.~\ref{fig:TCMPD}(e), $\alpha_{\rm min}\approx 0$ for the upper wedge \footnote{Note that the semi-infinite full-shell nanowire is strictly trivial for $\alpha=0$, but can be topological for vanishingly small SO coupling.}.

The topological regions also move to  smaller values of the chemical potential $\mu$ as $W$ increases. The reason is that increasing $W$ decreases the radial confinement energy $\langle p_r^2\rangle/2m^*$ in the Hamiltonian \eqref{solidrot}.
If we had plotted the phase diagrams against the Fermi energy $\tilde\mu$ [as we did in the hollow-core nanowire of Fig.~\ref{fig:HCA}(g)], they would cover a similar range of $\tilde\mu$ values\footnote{For $U(r)=0$, we have that $\tilde\mu\equiv\mu-\langle p_r^2\rangle/2m^*$, see Sec.~\ref{Sec:HCNw}. As $W\rightarrow 0$, i.e., in the hollow-core approximation, the radial confinement energy goes to infinity.}.

Note also that in the upper wedged-shaped topological regions of Fig.~\ref{fig:TCMPD}(a-e) there are several topological islands (as we saw also in the hollow-core model). Conspicuously, there are no topologically protected MZMs in the lower wedged-shaped topological regions. We could perform an equivalent study to Fig.~\ref{fig:TCMpos} of the LDOS behavior with tubular-core thickness, but for parameters in the lower topological regions. All the phenomenology would be qualitatively the same (not shown) except for the fact that the whole extent of the Majorana ZEP against flux would be covered by dispersing CdGM analogs crossing zero energy. Given that the SO coupling needed to enter a topological phase is larger (in absolute value), and that there is no true topological minigap, the negative-$\alpha$ tubular-core nanowire is not a good candidate to look for Majorana bound states.

It is interesting to realize that, as $W$ increases, the values of $\Gamma_{\rm S}$ needed to induce an equivalent proximity effect strongly increase, see Fig.~\ref{fig:TCMPD}(f-j) and note the logarithmic scale in the vertical axes. This is so because the effect of the superconductor on the core is exerted through the self energy \eqref{shelfenergy} at the core's boundary, $r=R$, while the charge wavefunction spreads to $r<R$ for finite $W$.

In the phase diagrams of Fig.~\ref{fig:TCMPD} we are considering values of $\mu$ that are relatively close to the semiconductor band bottom for each $W$. This is, we are considering the first topological region corresponding to the $m_r=0$ radial quantum number. If we increased $\mu$, we could enter another topological region corresponding to $m_r=1$ and so on. For small $W$, the topological regions for different $m_r$ are well separated in $\mu$, but as $W\rightarrow R$, they come closer and can even touch, as we will see when we analyze the solid-core nanowire.

In this section we have analyzed the tubular-core model focusing on a core radius $R=70$~nm. For completeness, we show an equivalent study to Fig.~\ref{fig:TCMpos} but for $R=30$~nm in Fig.~\ref{fig:TCMdestructive} of App.~\ref{Ap:destructive}. There, $|\alpha_{\rm min}|\approx 0$, see Fig.~\ref{fig:TCMdestructive}(a). Particularly, for $\alpha=10$~meV\,nm and $W  = 10$~nm, we get a topological minigap $E_{\rm g}= 20$~$\mu$eV at $\Phi= 0.99\Phi_0$, see Fig.~\ref{fig:TCMdestructive}(c). Reducing the nanowire radius introduces another important change. While for a large radius such as $R=70$~nm the superconductor shell exhibits the so-called non-destructive LP regime (with abrupt, first-order, lobe-to-lobe transitions), nanowires with small radii display a quite different, destructive LP phenomenology. This is illustrated in Fig.~\ref{fig:TCMdestructive} for an $R=30$~nm nanowire. The full LDOS evolves continuously between lobes, which are now narrower and are separated by gapless and spectrally smooth regions, where the order parameter is suppressed through second-order transitions.

\subsection{Modified hollow-core model}
\label{Sec:MHCNw}

\begin{figure}
   \includegraphics[width=\columnwidth]{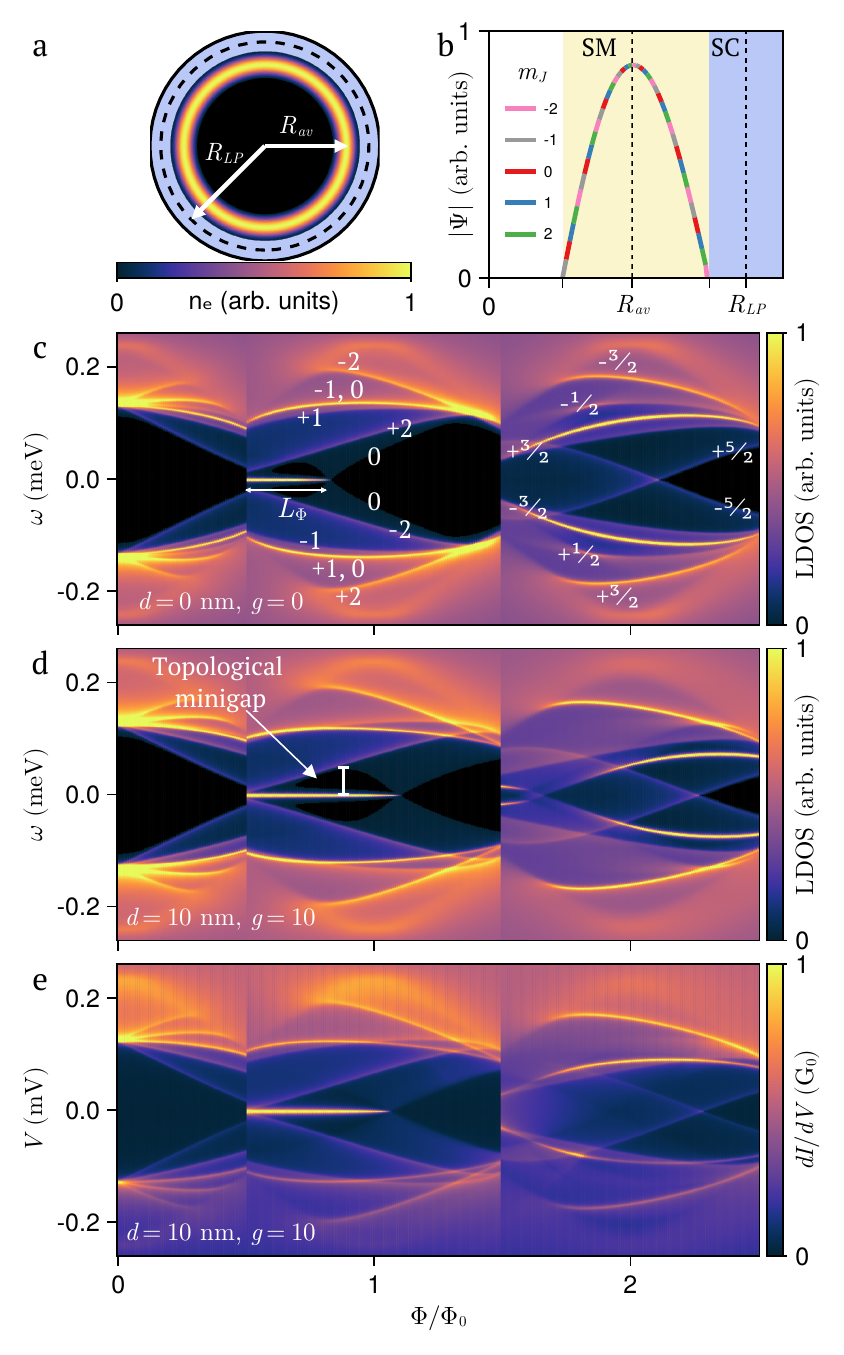}
   \caption{\textbf{Modified hollow-core model.} (a) Sketch of a tubular-core nanowire corresponding to Fig.~\ref{fig:TCMpos}(d), with $R=70$~nm, $W=20$~nm, $d=0$, $\alpha=50$~meV\,nm, $\mu=18.1$~meV and $\Gamma_{\rm S}=12\Delta_0$. The LP radius is $R_{\rm LP}=R+d/2$. Inside the core, simulation of the radial dependence of the electron density $n_e$, whose maximum value occurs at the average radius $R_{\rm av}$. (b) Wavefunction modulus of the populated $m_J$ subbands at $k_z=0$ in the normal state ($\Gamma_{\rm{N}}\rightarrow 0$) as a function of radial coordinate $r$. The wavefunction average radius is $R_\mathrm{av} = \langle r\rangle$. (c) Fitting of Fig.~\ref{fig:TCMpos}(d) using the modified hollow-core model with $R_\mathrm{av}=59.5$~nm extracted from (b), $\tilde\mu = 0.5$~meV and $\Gamma^{\rm av}_{\rm S} = 1.1\Delta_0$. The topological minigap is $E_{\rm g}=30$~$\mu$eV at $\Phi=0.71\Phi_0$. The topological transition happens at $\Phi=0.84\Phi_0$. The $m_J$ quantum numbers of the different GdGM analogs are shown in the $n=1,2$ lobes. $L_\Phi$ signals the MZM flux interval given in Eq.~\eqref{LphiMHC}. (d) Same as (c) but for a finite shell thickness $d=10$~nm and g-factor $g=10$. Now $E_{\rm g}=40$~$\mu$eV at $\Phi=0.84\Phi_0$. The topological transition happens at $\Phi=1.1\Phi_0$. (e) Differential conductance $dI/dV$ corresponding to (d) (in units of the conductance quantum $G_0$) versus normalized flux for a device like the one represented in Fig.~\ref{fig:sketch}(c) with tunnel potential barrier of height $10$~meV and length $50$~nm. Other parameters as in Fig.~\ref{fig:HCA}.}
   \label{fig:MHC}
\end{figure}

In this section we analyze an efficient approximation to calculate the full-shell nanowire phenomenology for a tubular-core model. This approximation is valid for semiconductor thicknesses $W$ for which the degeneracy points of the first lobe are well defined (approximately until they exit the lobe). For example, for the parameters of Fig.~\ref{fig:TCMpos}, it is valid for $W\lesssim 30$~nm. In this case, all the populated $m_J$ subbands have essentially the same radial wavefunction profile and hence the same $R_{\rm av}$, as explained in the previous section. We call this approximation the $\textit{modified}$ hollow-core model because the Hamiltonian is like the one of the hollow-core model but, instead of taking $r=R$ in Eq.~\eqref{solidrot}, we take $r=R_{\rm av}$ (while keeping $R_{\rm LP}$ unchanged). As in the case of the hollow-core model, the modified hollow-core Hamiltonian depends on $\tilde\mu$. However, now the self energy of Eq.~\eqref{shelfenergy} is evaluated at $r=R_{\rm av}$, and we denote the effective decay rate into the superconductor by $\Gamma^{\rm av}_{\rm S}$. This approximation was introduced in Ref.~\onlinecite{Paya:PRB23} for the case of $\alpha=0$, and is extended here for arbitrary $\alpha$.

In Fig.~\ref{fig:MHC} we concentrate on the case of a tubular-core nanowire with $R=70$~nm and $W=20$~nm, i.e., like the one in Fig.~\ref{fig:TCMpos}(d). Note that for this tube thickness, the $\pm\omega$ degeneracy points are still visible at the right edge of the $n=1$ lobe and that the Majorana ZEP covers approximately one third of the lobe flux interval. The local electronic density, defined as the integral over all energies of the LDOS times the Fermi distribution, $n_e=\int_{-\infty}^{\infty}\rho(\omega)f(\omega)$, is plotted in the sketch of Fig.~\ref{fig:MHC}(a). It is maximum at $r=R_{\rm av}$ and decays towards the semiconductor tube boundaries. Figure~\ref{fig:MHC}(b) shows the wavefunction modulus of all the populated $m_J$ subbbands in the normal state ($\Gamma_{\rm S}\rightarrow 0$) for a chemical potential of $\mu=18.1$~meV. They all have the same radial profile, which is maximum at $R_{\rm av}$ (approximately at the center of the tube thickness) and decays to zero at the tube inner and outer radii. If a finite $\Gamma_{\rm S}$ had been chosen for this plot, the wavefunction at $r=R$ would have a finite small value, as a result of the leakage of the core's charge density into the superconductor.

Now we fit the LDOS of Fig.~\ref{fig:TCMpos}(d) with the modified hollow-core model. We fix the average radius found in Fig.~\ref{fig:MHC}(b), $R_{\rm av}=59.5$~nm, and adjust the Fermi energy $\tilde\mu$ and effective decay rate $\Gamma^{\rm av}_{\rm S}$ as fitting model parameters \footnote{Note that the decay rate $\Gamma_{\rm S}$ in the hollow-core approximation is in general smaller than $\Gamma^{\rm av}_{\rm S}$ in the tubular-core model. The reason is that, in the modified hollow-core model, the self energy of Eq.~\eqref{shelfenergy} is evaluated at $r=R_{\rm av}$ where all the charge density is located. However, in the tubular-core model, the self energy is evaluated at the core boundary, $r=R$, whereas the charge density spreads throughout the tube thickness $W$.}. The result of the LDOS fit for $d=0$ is shown in
Fig.~\ref{fig:MHC}(c), which is almost indistinguishable from Fig.~\ref{fig:TCMpos}(d), while the fitted topological phase boundary is shown in orange in Fig.~\ref{fig:TCMPD}(c,h). The accuracy of both fits attests to the validity of the modified hollow-core approximation.
A maximum topological minigap of $E_{\rm g}=30$~$\mu$m is obtained at $\Phi=0.71\Phi_0$, see Fig.~\ref{fig:MHC}(c), exactly as in Fig.~\ref{fig:TCMpos}(d). The generalized angular momentum quantum numbers $m_J$ of the different subgap states are also shown in the $n=1,2$ lobes. Note that CdGM analogs with positive slope have positive $m_J$ and vice versa. Focusing on states with positive slope for $\omega>0$ and negative slope for $\omega<0$, the Van Hove singularities with larger $|m_J|$ tend to have smaller energies in absolute value in all lobes. The exception is the $m_J=0$ mode in the topological regime that, as we explained in Sec.~\ref{Sec:HCNw}, evolves strongly with $\alpha$ towards zero energy to carry out the topological transition with a gap closing and reopening at a finite flux. The rest of the CdGM analogs are essentially unaffected by the SO coupling $\alpha$.

\begin{figure}
   \includegraphics[width=\columnwidth]{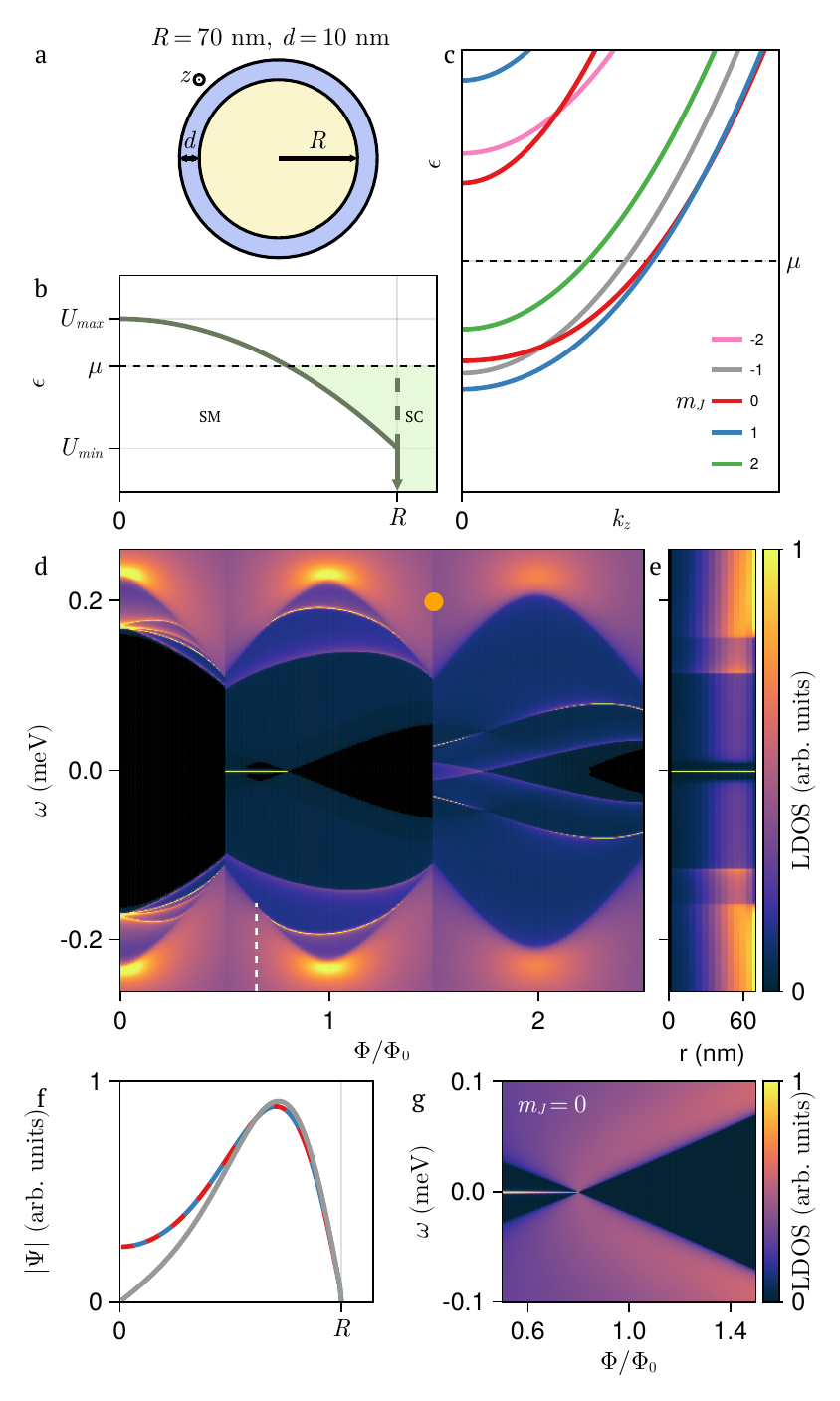}
   \caption{\textbf{Solid-core model, lowest radial mode.} (a) Sketch of the solid-core nanowire with $R=70$~nm and $d=10$~nm. (b) Electrostatic potential energy displaying a dome-shaped radial profile inside the semiconductor core, with $U_{\rm max} = 0$~meV, $U_{\rm min} = -30$~meV and $\mu = -11$~meV. The chemical potential is such that only the lowest radial subband $m_r=0$ is populated. (c) Band structure in the normal state ($\Gamma_{\rm S}\rightarrow 0$) as a function of the longitudinal wave vector $k_z$. The number of occupied subbands depends on $\mu$. Different colors signal different values of the quantum number $m_J$ in the $n=1$ LP lobe. The average SO coupling is $\langle \alpha \rangle = 20$~meV\,nm [corresponding to a prefactor $\alpha_0 = 46.7$~nm$^2$ in Eq.~\eqref{Eq:SOC}]. (d) LDOS at the end of a semi-infinite solid-core nanowire (in arbitrary units) as a function of energy $\omega$ and applied normalized flux $\Phi/\Phi_0$. There is a topological minigap $E_{\rm g}=15$~$\mu$eV at $\Phi=0.65\Phi_0$ (marked with a dashed white line). The parameters of the wire correspond to the pink dot in the topological phase diagram of Fig.~\ref{fig:SCMPD}(c). (e) Radial dependence of the previous LDOS at that flux. (f) Wavefunction modulus of the populated subbands at $k_z=0$ in the normal state ($\Gamma_{\rm{N}}\rightarrow 0$) as a function of radial coordinate $r$. (g) LDOS in the $n=1$ lobe coming only from the $m_J=0$ subband. For panels (c, f), $\Phi=0.65\Phi_0$. Parameters: $\Gamma_{\rm{S}}=40\Delta_0$, $g=10$. Other parameters as in Fig.~\ref{fig:HCA}.}
   \label{fig:SCMposmr0}
\end{figure}

\begin{figure}
   \centering
   \includegraphics[width=\columnwidth]{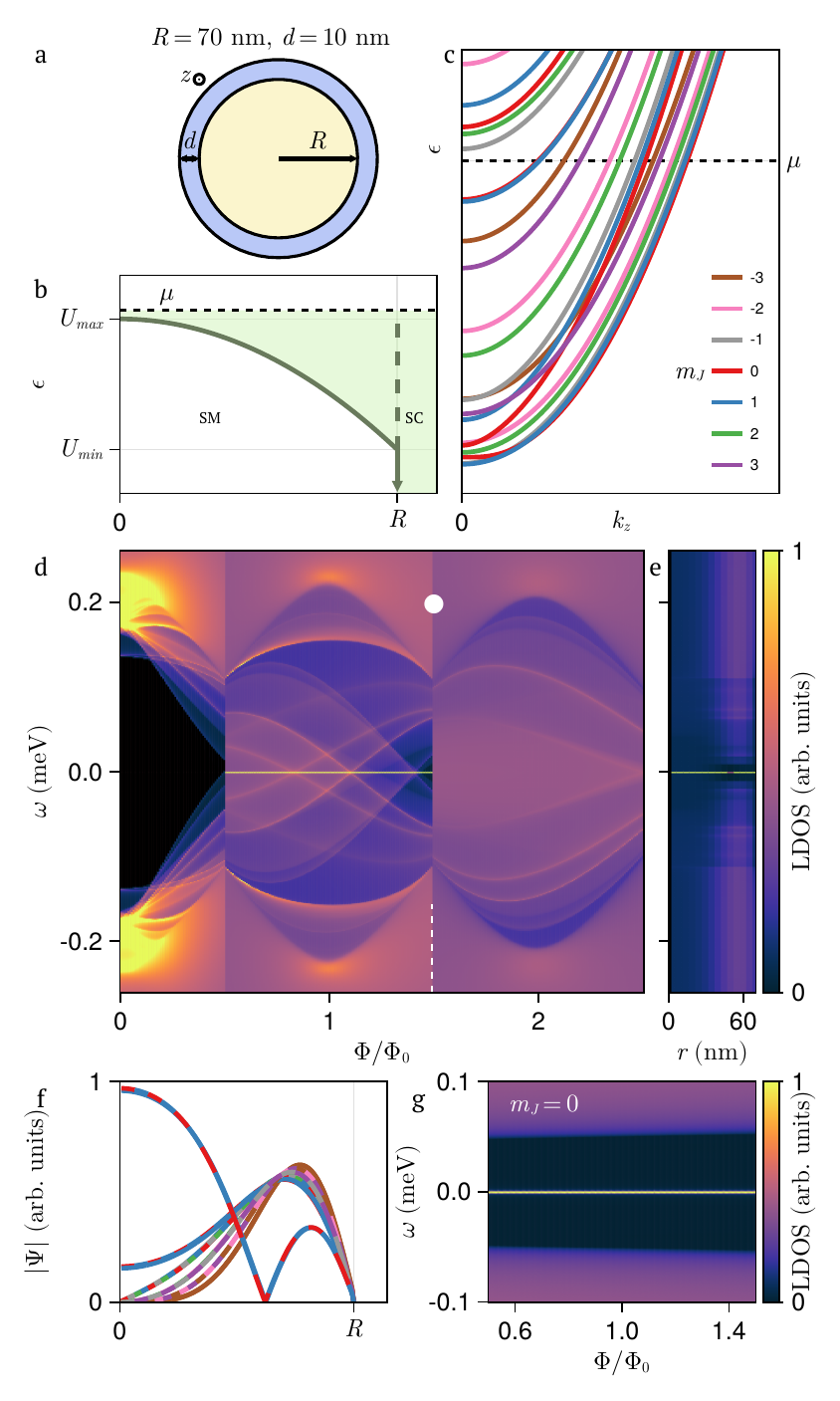}
   \caption{\textbf{Solid-core model, first radial mode.} Same as Fig.~\ref{fig:SCMposmr0} but for a positive chemical potential, $\mu = 2$~meV, such that not only the lowest radial subband $m_r=0$ is populated, but a few $m_J$ subbands have $m_r=1$. There is no topological minigap in this case for any flux. The parameters of the wire correspond to the white dot in the topological phase diagram of Fig.~\ref{fig:SCMPD}(b). For panels (c, e, f), $\Phi = 1.49 \Phi_0$.}
   \label{fig:SCMposmr1}
\end{figure}

The modified hollow-core model also allows us to derive an analytical approximation for the MZM flux interval $L_\Phi$ for a tubular-core nanowire, again as long as we have well defined degeneracy points in the first lobe. Its expression is the same as Eq.~\eqref{Lphi} but replacing $R$ by $R_{\rm av}$,
\begin{equation}
\label{LphiMHC}
    L_\Phi = \mathrm{clip}\left(\frac{\Phi^{(2)}_{\rm TT}(R=R_{\rm av})}{\Phi_0} - \frac{1}{2}, [0, 1]\right).
\end{equation}
By inspecting this equation, it is then possible to understand why the left MZM flux interval in Fig.~\ref{fig:TCMpos} grows as $W$ increases in each panel. This behavior is essentially dominated by the factor $(R_{\rm LP}/R_{\rm av})^2$ in $\Phi^{(2)}_{\rm TT}(R=R_{\rm av})$, see Eq.~\eqref{PhiTT}, which increases as $R_{\rm av}$ is reduced.

Lastly, in Fig.~\ref{fig:MHC}(e) we show the differential tunneling conductance $dI/dV$ for the semi-infinite tubular-core of Fig.~\ref{fig:MHC}(d), computed following App.~\ref{Ap:Obs}. The schematics of the tunneling spectroscopy device can be seen in Fig.~\ref{fig:sketch}(c), with a $z$-dependent tunnel barrier of height $10$~meV and length $50$~nm. We observe that the $dI/dV$ is a faithful measurement of the LDOS at the end of the full-shell wire, specially in the $n=0$ and $n=1$ LP lobes. A more detailed discussion on the relation between LDOS and $dI/dV$ in this type of wires can be found in Ref.~\onlinecite{Paya:PRB23}.

\subsection{Solid-core model}
\label{Sec:SCNw}

We finally present results for the solid-core model for the full-shell nanowire [Fig.~\ref{fig:SCMposmr0}(a)]. The model includes a dome-like electrostatic potential profile [Fig.~\ref{fig:SCMposmr0}(b)] and the associated SO coupling inside the core, which is otherwise no longer hollow. In this case we consider parameters such that there is a radially averaged $\langle \alpha \rangle = 20$~meV\,nm. All the details of the Hamiltonian and the calculation of the spatially-varying SO coupling can be found in App.~\ref{Ap:Model}.

We first turn our attention to a situation where only the lowest radial subband $m_r=0$ is occupied, to make connection with Sec. \ref{Sec:TCNw}. This happens when the Fermi level is sufficiently below the top of the dome, see Fig.~\ref{fig:SCMposmr0}(b). Even though $m_r=0$, there can be several filled angular momentum subbands $m_J$ depending on the value of $\mu$. In a topological phase, like the one considered in Fig.~\ref{fig:SCMposmr0}(c), $\mu$ lies between the two subbands of the $m_J=0$ pair (in red) \footnote{These two red subbands with $m_J=m_r=0$ have quantum numbers $(m_l=-1,m_s=1/2,m_n=1/2)$ for the top one and $(m_l=0,m_s=-1/2,m_n=1/2)$ for the bottom one, see App.~\ref{Ap:quantumnumbers} for a discussion of quantum numbers.}. The LDOS at the end of the wire is plotted versus flux in Fig.~\ref{fig:SCMposmr0}(d). We observe that the qualitative behavior is quite similar to that of the tubular-core model: skewed CdGM analogs, shifted gap, and Majorana ZEP covering a finite flux interval on the left side of the $n=1$ LP lobe. A topological minigap with a maximum of $15$~$\mu$eV is reached at $\Phi=0.65\Phi_0$. At this same field, Fig.~\ref{fig:SCMposmr0}(e) shows the LDOS spatially resolved in radial coordinate $r$, which confirms that all modes belong to the $m_r=0$ sector (they have no radial nodes, including the MZM). We observe two differences with respect to the tubular case. First, the degeneracy point is no longer visible, see Fig.~\ref{fig:SCMposmr0}(d). This is a result of the reduction of $R_\mathrm{av}$, which is apparent from Fig.~\ref{fig:SCMposmr0}(f). Secondly, this figure also shows that, while wavefunctions with different $m_J$ are still quite similar to one another, they now spread all the way to $r=0$. Their asymptotic behavior is $|\Psi|\sim r^{m_l}$, so that those with $m_l=0$ have a finite value at $r=0$. In Fig.~\ref{fig:SCMposmr0}(g) we show the LDOS of only the $m_J=0$ subband.

\begin{figure}
   \includegraphics[width=\columnwidth]{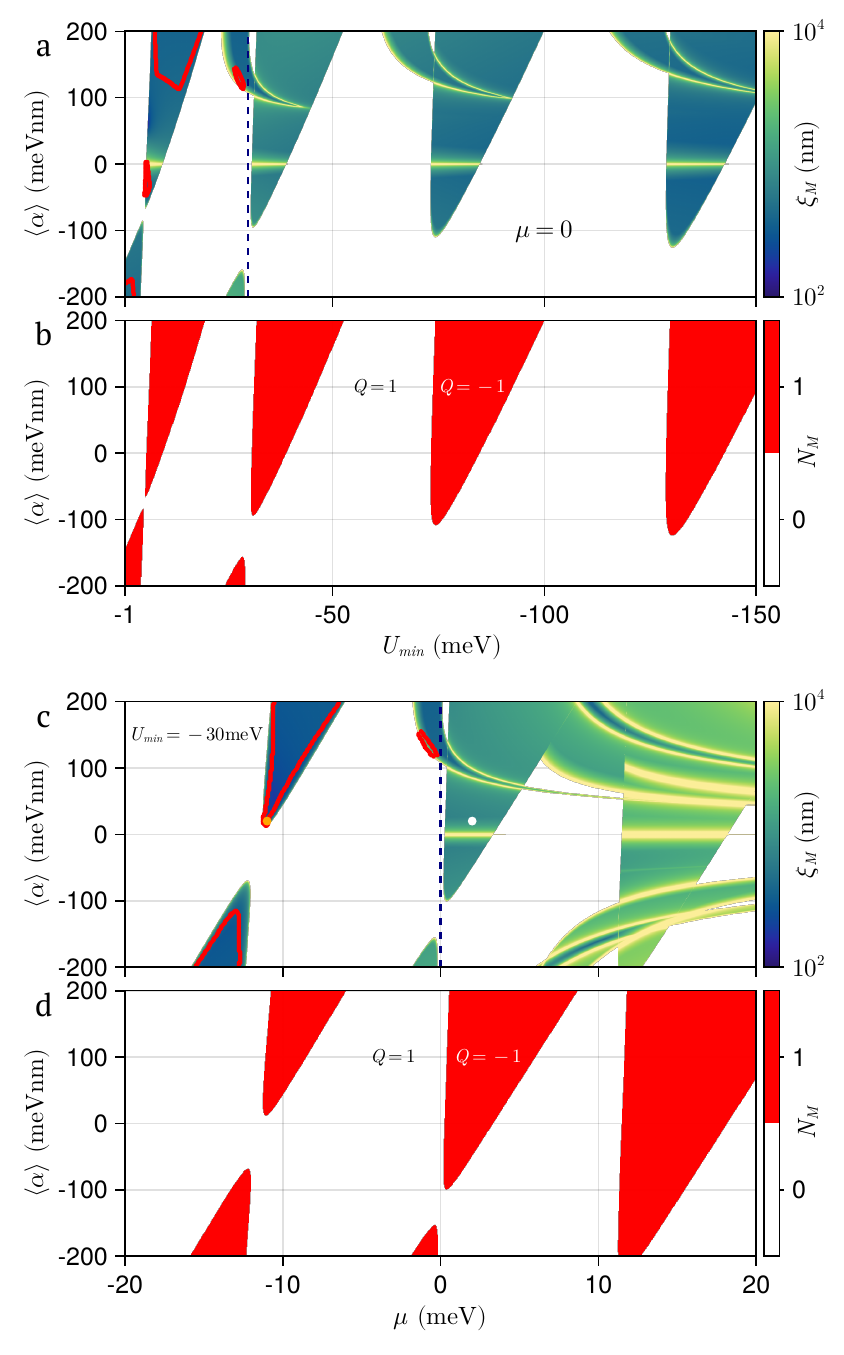}
\caption{\textbf{Topological phase diagram for the solid-core model}. Map of ZEPs (a) and topological phase diagram (b) as a function of radially averaged SO coupling $\langle\alpha\rangle$ and potential profile minimum $U_{\rm min}$, with $U_{\rm max}=0$ and $\mu = 0$~meV. The color scale in (a) represents the Majorana localization length $\xi_{\rm M}$ at $\Phi = 0.51 \Phi_0$. The color bar in (b) represents the number of MZMs $N_{\rm M}$, related to the topological invariant by $\mathcal{Q}=(-1)^{N_{\rm M}}$ (red is non-trivial). (c, d) Same as (a, b) but as a function of a rigid shift of the chemical potential $\mu$, fixing $U_{\rm min} = -30$~meV. Colored dots mark parameter values for Figs.~\ref{fig:SCMposmr0} (orange) and \ref{fig:SCMposmr1} (white). Other parameters: $R=70$~nm, $d=10$~nm, $\Delta_0 = 0.23$~meV, $\xi_{\rm d} = 70$~nm, $\Gamma_{\rm S} = 40\Delta_0$, $g = 10$ and $a_0=5$~nm.}
   \label{fig:SCMPD}
\end{figure}

A different prototypical scenario is obtained when the chemical potential lies close to or above the top of the dome-like electrostatic profile, see Fig.~\ref{fig:SCMposmr1}(b), so that more radial momentum subbads become populated. We consider in particular the case in which only up to the $m_r=1$ subband is filled for the lowest $m_J$ sector, see the band structure of Fig.~\ref{fig:SCMposmr1}(c). There we show a situation with the Fermi level between the two $m_J=0$ (red) subbands in the $m_r=1$ sector. When proximitized by an $n=1$ fluxoid, this configuration results in a $m_r=1$ MZM. The other two $m_J=0$ (red) subbands with $m_r=0$ have much lower energy, close to the conduction band bottom, and as they are both filled, they do not contribute to create Majoranas.

The LDOS versus flux and radial coordinate are depicted in Fig.~\ref{fig:SCMposmr1}(d,e), respectively. Now the Majorana ZEP extends throughout the first lobe, like in the tubular-core model for a sufficiently large thickness $W$. Moreover, there are many CdGM analogs dispersing with flux and crossing zero energy, coming from all the populated $m_J$ subbands. They cover the whole lobe width (and height), generating a dense LDOS background with which the ZEP coexists.

The wavefunctions of the different subbands can be seen in Fig.~\ref{fig:SCMposmr1}(f). Now there are two types of wavefunctions. Those corresponding to $m_r=0$, which have only one maximum as a function or $r$ at a similar radius for all $m_J$. Then we have those corresponding to $m_r=1$, which have two local maxima separated by one node at a finite $r$. Since the MZM comes from the $m_r=1$, $m_J=0$ subband, it also exhibits this type of radial profile, with a single node at a finite $r$ and the first maximum at $r=0$, see \ref{fig:SCMposmr1}(e). Qualitatively, this results in a strong reduction of the average radius $R_{\rm av}$, which following the arguments given above about the degeneracy point position, strongly shifts the corresponding $m_r=1$ CdGM analogs towards the right. Consequently, the whole first lobe becomes gapless, and the Majorana ZEP extends all across it.

The topological phase diagram of the solid-core nanowire is shown in Fig.~\ref{fig:SCMPD}. We plot it as a function of the radially averaged SO coupling $\langle\alpha\rangle$, both for positive and negative values, and as a function of the potential profile minimum $U_{\rm min}$ [in Fig.~\ref{fig:SCMPD}(a,b)] or the chemical potential $\mu$ [in Fig.~\ref{fig:SCMPD}(c,d)]. In Fig.~\ref{fig:SCMPD}(a,c) the color scale represents the Majorana localization length $\xi_{\rm M}$ of the ZEP at $\Phi=0.51\Phi_0$, while in Fig.~\ref{fig:SCMPD}(b,d) we represent the  number $N_{\rm M}$ of Majorana zero modes, given in terms of the topological invariant $\mathcal{Q}$ by
\begin{equation}
\mathcal{Q} = (-1)^{N_{\rm M}} = \mathrm{sign}\,\prod_{m_J}\mathrm{Pf}\left[\sigma_y\tau_y H(k_z=0, m_J)\right],
\end{equation}
where Pf is the Pfaffian, see App.~\ref{Ap:pwave} and Refs.~\onlinecite{Vaitiekenas:S20} and  \onlinecite{Ghosh:PRB10}.

It is possible to see how the different topological regions for each $m_r$ come into play as $|U_{\rm min}|$ increases (and thus the wire doping), starting from $m_r=0$ to the left of the diagram. Each $m_r$ introduces its own wedge-shaped topological region. Notably, the $R=70$~nm solid-core nanowire can be in the topological ($\mathcal{Q}=-1$) phase for small SO coupling $\langle\alpha\rangle\approx 0$ (note that at $\langle\alpha\rangle=0$ the system is strictly trivial). The negative-$\langle\alpha\rangle$ topological regions appear only for very large values of $|\langle\alpha\rangle|$, a tendency that increases with doping. The LDOS also exhibits a ZEP inside $L$-shaped non-topological regions between the wedges at very large values of $\alpha$. Since this regions have $\mathcal{Q}=1$, the associated zero mode is not a Majorana of topological origin. We checked numerically, however, that the ZEP seems to be a robust feature, and that it arises from a band inversion at a finite $k_z$. We leave a more detailed analysis of its nature to a future work.

Conspicuously, topologically protected parameter islands (enclosed by a red curve) in the phase diagrams only happen for the $m_r=0$ wedges, both upper and lower, see Fig.~\ref{fig:SCMPD}(a,c). Thus, for a cylindrical full-shell solid-core nanowire there can be protected MZMs only when it behaves approximately like a tubular-core one. This might be impossible in practice, in view of the shape of the electrostatic potential profiles calculated self-consistently using a Schrödinger-Poisson approach in previous works \cite{Vaitiekenas:S20, Woods:PRB19}. For pristine solid-core nanowires with more than one occupied radial mode, the proliferation of CdGM analogs prevents the appearance of topological minigaps.

\section{Mode-mixing perturbations}
\label{Sec:modemixing}

\begin{figure}
   \includegraphics[width=\columnwidth]{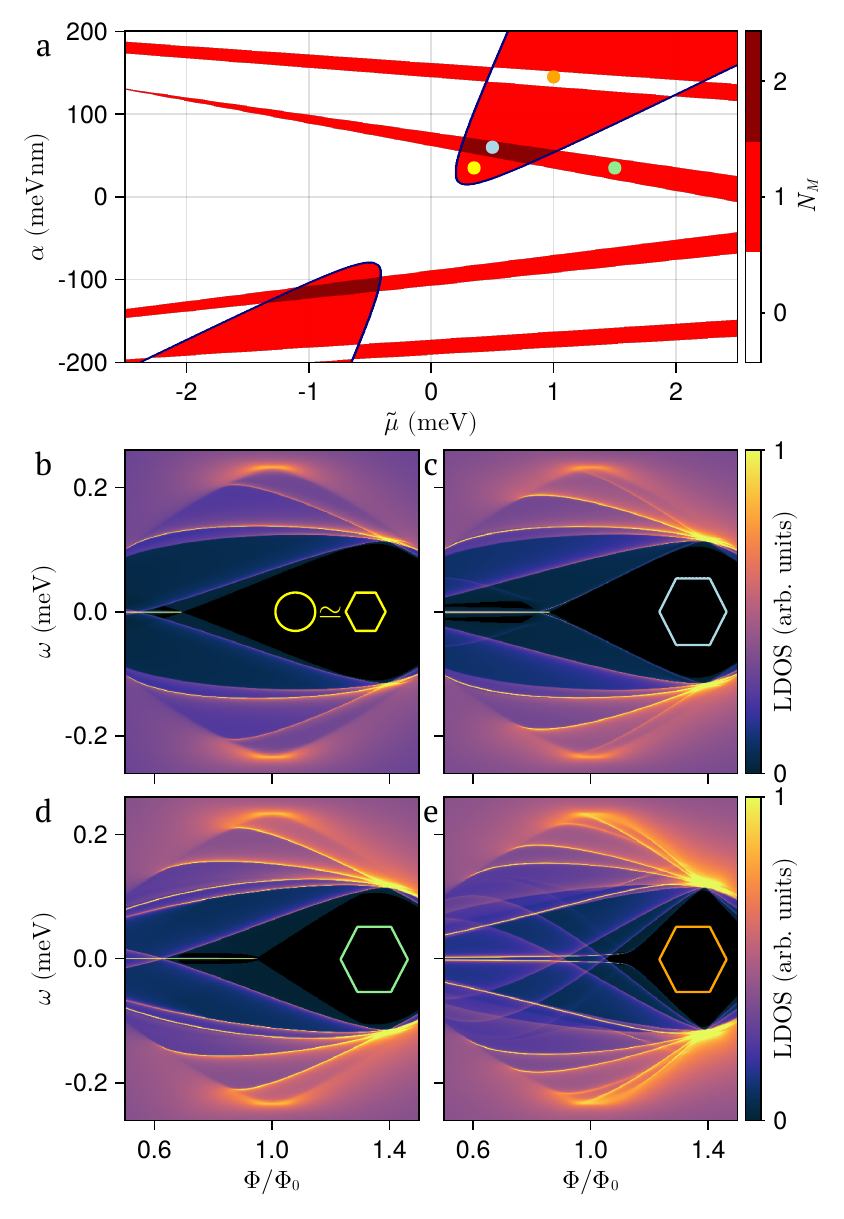}
    \caption{\textbf{Hexagonal cross-section tubular-core model.} (a) Topological phase diagram as a function of SO coupling $\alpha$ and Fermi energy $\tilde{\mu}$. The color bar represents the number of MZMs $N_{\rm M}$. (b-e) LDOS at the end of a semi-infinite nanowire as a function of $\omega$ and $\Phi/\Phi_0$ in the $n=1$ LP lobe for the four colored points in the phase diagram of (a) (coded by the color of the cross section profile). (b) is indistinguishable from a cylindrical case with the same parameters. (c) exhibits two non-interacting MZMs, so their ZEPs add up, while in (e) they hybridize and their ZEPs split. (d) has a single MZM arising from a $m_J = \pm 3$ mode mixing. Parameters: (b) $\tilde{\mu} = 0.35$~meV, $\alpha = 35$~meV\,nm, $E_{\rm g} = 10$~$\mu$eV, (c) $\tilde{\mu} = 0.5$~meV, $\alpha = 60$~meV\,nm, $E_{\rm g} = 20$~$\mu$eV, (d) $\tilde{\mu} = 1.5$~meV, $\alpha = 35$~meV\,nm, $E_{\rm g} = 5$~$\mu$eV, (e) $\tilde{\mu} = 1$~meV, $\alpha = 145$~meV\,nm. Other parameters as in Fig.~\ref{fig:MHC}(c).}
   \label{fig:Hex}
\end{figure}

\begin{figure*}
\centering
\includegraphics[width=\textwidth]{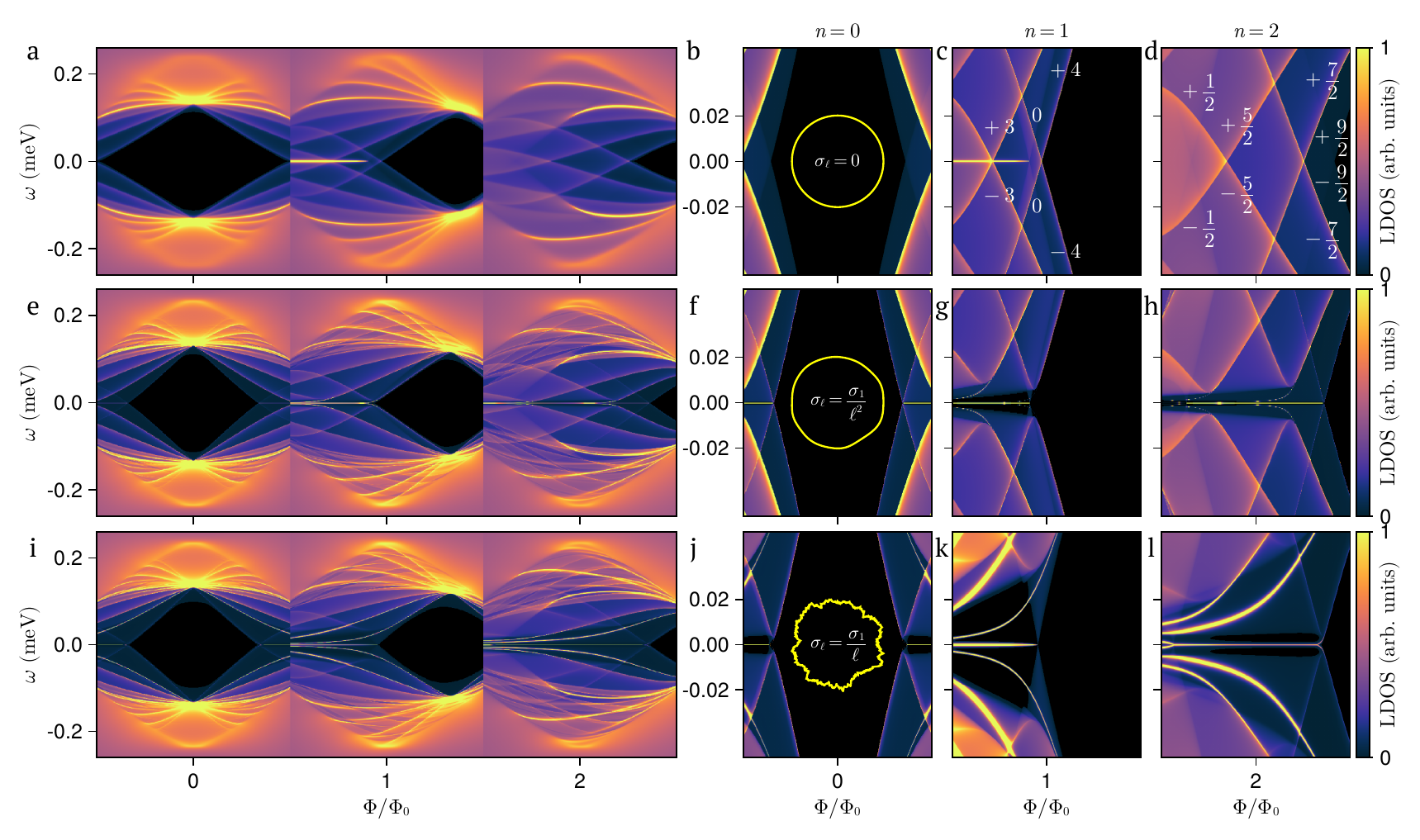}
\caption{\textbf{Mode mixing in the modified hollow-core model.}
    First row: (a) Equivalent LDOS to Fig.~\ref{fig:MHC}(c) but for a SO coupling $\alpha = 100$~meV\,nm, so that the left ZEP of the $n=1$ LP lobe is covered by several CdGM analogs. (b-d) Blow ups around zero energy $\omega$ of the $n=0$ (b), $n=1$ (c) and $n=2$ (d) LP lobes. The $n=1$ Majorana ZEP is visible in (c). There is no topological minigap for any flux. The quantum numbers $m_J$ are highlighted in (b) and (c).
    Second row: Same as first row but for a modified hollow-core nanowire subject to a mode mixing of $\sigma /R_0 = 0.2$, using a smooth disorder model in Eq.~\eqref{harmonics}. The disorder-induced wavefunction distortion generates ZEPs at the edges of the $n = 0$ lobe, see (f), and a long ZEP at the left side of the $n = 2$ LP lobe, see (h). There is a small topological minigap in the three lobes. Third row: Same as second row but with a non-smooth disorder model representing atomic-sized defects. The ZEPs are practically the same but all topological minigaps are significantly larger. Other parameters as in Fig.~\ref{fig:MHC}(c).
}
\label{fig:modemixing}
\end{figure*}

\begin{figure}
   \includegraphics[width=\columnwidth]{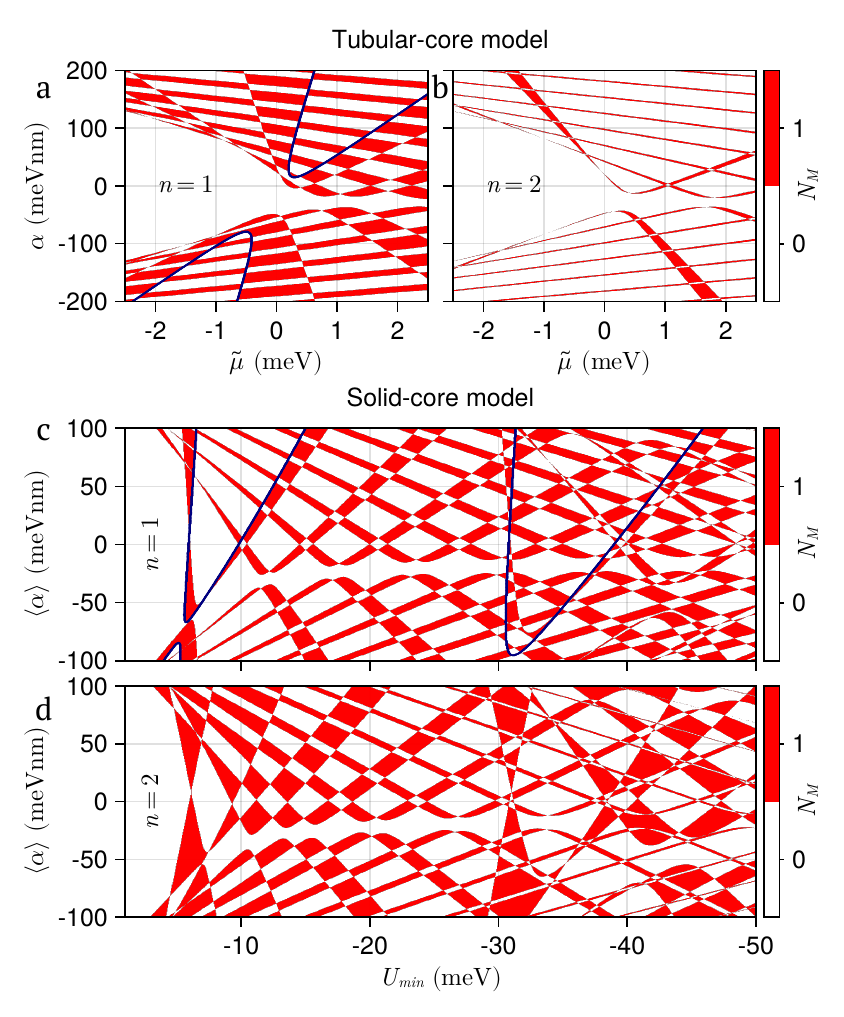}
   \caption{\textbf{Topological phase diagrams in the presence of generic mode mixing}. (a) Topological phase diagram as a function of SO coupling $\alpha$ and Fermi energy $\tilde{\mu}$ for the tubular-core nanowire of Fig.~\ref{fig:TCMPD}(a, b) in the $n=1$ LP lobe (at $\Phi = 0.51 \Phi_0$). The color bar represents the number of MZMs $N_{\rm M}$. The boundary of the $m_J = 0$ topological region without mode mixing is contoured with a blue line. (b) Same as (a) but for the $n=2$ LP lobe (at $\Phi = 1.51 \Phi_0$).  (c, d) Same as (a, b), respectively, but for the solid-core nanowire of Fig.~\ref{fig:SCMPD}(b). These phase diagrams are given as a function of radially averaged SO coupling $\langle\alpha\rangle$ and
   potential profile minimum $U_{\rm min}$, with $U_{\rm max}=0$ and $\mu = 0$~meV.}
   \label{fig:MMPD}
\end{figure}

So far we have assumed nanowires with a perfectly circular cross section, and hence, with independent $m_J$ modes. In this section we would like to understand the effect of non-cylindrical symmetry on the LDOS. This includes both the typical hexagonal shape often adopted by crystalline nanowires, and more general breaking of cylindrical symmetry as produced e.g. by disorder. Both introduce an essential new ingredient in the form of mode mixing between modes of different $m_J$.

Here we assume cross section distortions to be $z$-independent. This is done for simplicity, since we are focusing on the mode-mixing effect. It is known from the literature of partial-shell nanowires that $z$-dependent disorder along the length of the wire is common and, when strong enough, it is be very detrimental to the survival and protection of MZMs~\cite{Kells:PRB12,Prada:NRP20}. Strong impurities break the wire down into several effectively distinct sections, creating a string of split Majoranas that take the form of trivial, low-energy subgap states\cite{Stanescu:PRB13, Fleckenstein:PRB18}. In addition, smooth potential profiles along $z$ may create quasi-Majoranas\cite{Kells:PRB12, Prada:PRB12, Penaranda:PRB18, Moore:PRB18, Vuik:SP19}. We expect this type of disorder to have similar effects in full-shell geometries.

Modelling arbitrary cross-section deformations or cross-section disorder exactly is computationally expensive, specially for large cross-section areas. Instead, we consider here a phenomenological model that can help us understand the consequences of mode-mixing perturbations transparently and at a reduced computational cost. It is based on the modified hollow-core approximation, generalized to include radial shifts of the wavefunction position  so that $R_\mathrm{av}$ depends on the polar angle $\varphi$ around the wire axis. In general, these $\varphi$-dependent shifts will couple modes with $m_J\neq m_J'$ with a strength related to the $m_J-m_J'$ Fourier harmonic of $R_\mathrm{av}(\varphi)$.
We consider three possibilities for $R_\mathrm{av}(\varphi)$: Hexagonal cross sections, random but smooth geometric distortions of the cross section, and non-smooth distortions representing random atomic-size impurities. The last two cases represent nanowires with disorder, typically on the shell surface or at the core-shell interface. The amount of mode-mixing can be quantified by a dimensionless parameter $\sigma/R_0$, where $\sigma$ is the standard deviation of $R_\mathrm{av}$ around its average value $R_0$. Full details of the method are given in App.~\ref{Ap:modemixing}.

Let us first compare clean nanowires with circular and hexagonal cross sections. In Fig.~\ref{fig:Hex} we show the topological phase diagram and the LDOS for a hexagonal tubular-core nanowire with $R=70$~nm and $W=20$~nm (these are now defined as $\varphi$-averages). In Fig.~\ref{fig:Hex}(b) the LDOS is evaluated for the parameters of the yellow dot in Fig.~\ref{fig:Hex}(a). This LDOS is essentially indistinguishable from that of a circular cross-section with the same $R$ and $W$. Rather surprisingly, this is also true for all the other parameters colored in red within the wedged-shaped topological regions (contoured by a blue curve). These wedged-shaped regions are in fact the same as the topological phase boundaries for a cylindrical nanowire with the same $R$ and $W$.

The hexagonal angular profile $R_\mathrm{av}(\varphi)$ of the core wavefunction position \footnote{Note that for the wavefunction to have an hexagonal shape in a hexagonal cross-section nanowire, this must be rather close to the core-shell interface. Otherwise, the average radius tends to develop a circular shape quickly as the wavefunction moves away from the boundary and towards the cylinder axis, further improving the cylindrical approximation for pristine nanowires.} introduces, nevertheless, some additional features in the topological phase diagram, Fig. ~\ref{fig:Hex}(a). There are new, narrow topological stripes that correspond to regions with an odd occupancy of $m_J=\pm 3$, $m_J=\pm 6$ (shown) and more generally $m_J=\pm 3N$ subbands, where $N$ is a positive integer.
Within the stripes, new MZMs appear. If they coincide with existing $m_J=0$ MZMs, the two MZMs may split, yielding pairs of trivial near-zero modes (white regions in the intersection of the stripes and the $m_J=0$ hyperbolas). This only happens for $m_J=\pm 6N$, since the finite $m_J$ should mode-mix with $m_J=0$ to produce the splitting. Otherwise, the splitting is zero, and two orthogonal MZMs coexist at the end of the nanowire (darker red regions).
The LDOS in these new regions of the phase diagram are plotted in Fig. ~\ref{fig:Hex}(c-e) for the different colored dots in Fig. ~\ref{fig:Hex}(a). The effect of mode-mixing is most visible around zero energy. Away from zero energy, the LDOS is again essentially indistinguishable from the circular cross-section case.

The mechanism behind the formation of the new MZMs is quite unexpected. In the absence of mode mixing, modes with finite $m_J$ become populated when they first cross the Fermi level. This happens each time we enter or exit a stripe. Due to the particle-hole symmetry relating $m_J$ and $-m_J$ sectors, $\pm m_J$ mode partners disperse as $\epsilon \sim \pm k_z^2$ at the crossing. The effect of a finite $m_J\leftrightarrow -m_J$ mode-mixing on this parabolic touching point is linear in $k_z$ (see App.~\ref{Ap:pwave}), turning the parabola crossing into a linear-in-$k_z$ band inversion. Mode mixing, therefore, takes exactly the form of a $p$-wave pairing between particle-hole partners. The magnitude of the pairing is proportional to the $\delta R_{2m_J}$ Fourier harmonic, which explains why in a pristine hexagonal nanowire, the new topological regions arise only in odd lobes for odd occupation of modes with $m_J=\pm 3N$, since the hexagon has only nonzero $\delta R_{6N}$ harmonics. By the same token, the splitting of the associated MZM with a pre-existing $m_J=0$ MZM is only possible if their $m_J$ difference is $6N$. Otherwise the two MZM remain decoupled. In all cases, the new MZMs induced by the hexagonal cross section have a tiny topological minigap of around 1~$\mu$eV, given the small magnitude of $\delta R_{6}\approx 0.05R_0$. Note also that in even lobes (not shown), the hexagonal section does not introduce any new MZM, or any significant change in the finite energy LDOS, since $2m_J$ is then always odd.

The above analysis is naturally extended to the case of disorder-induced mode mixing. In this case all harmonics $\delta R_{\ell}$ are present. The difference between smooth and non-smooth disorder is their decay with harmonic index $\ell$, see App.~\ref{Ap:modemixing}. The corresponding LDOS results are shown in the second and third rows of Fig.~\ref{fig:modemixing}, respectively, to be compared to the cylindrical case in the first row. The corresponding $R_\mathrm{av}(\varphi)$ profiles are shown in yellow in Figs.~\ref{fig:modemixing}(b, f, j). Small-energy blowups of the $n=0$, $n=1$ and $n=2$ lobes are shown in the second to fourth columns for each case. Numbers in Fig.~\ref{fig:modemixing}(c,d) denote the $m_J$ of each CdGM analog without $m_J$ mixing.  We have chosen the parameters $R=70$~nm, $W=20$~nm and $d=0$, as in Fig.~\ref{fig:MHC}(c), but now we have increased the SO coupling to $\alpha=100$~meV\,nm. This stronger $\alpha$ takes us out of the topologically protected island in the perfectly cylindrical case, so that while a MZM is still present throughout half of the first lobe, it coexists with a collection of CdGM analogs crossing at zero energy, see Fig.~\ref{fig:modemixing}(c).
Without mode-mixing, therefore, there is no MZM minigap in the first lobe.

Introducing a mode-mixing distortion with a small $\sigma/R_0 = 0.2$ has two distinct effects. First, for $m_J$ sectors with even occupation, a trivial minigap is opened around zero energy in the CdGM analogs, so that any preexisting $m_J=0$ MZM will no longer overlap with them and will become topologically protected, see Fig.~\ref{fig:modemixing}(g,k). However, if any $m_J$ sector has an odd number of occupied modes, the minigap opened by the $\delta R_{2m_J}$ harmonic will be topological, introducing a new MZM. This mechanism of Majorana formation generalizes the simpler $m_J=0$ mechanism that operates in odd lobes in the absence of mode mixing, see App.~\ref{Ap:HCA_Topology}. Remarkably, the new MZMs may also form in even lobes, since finite harmonics with odd $\ell$ are now present, unlike in pristine hexagonal nanowires. Figures~\ref{fig:modemixing}(f,j) and \ref{fig:modemixing}(h,l) showcase finite $m_J$ MZMs in the $n=0$ and $n=2$ lobe, respectively. Moreover, the new MZMs will always hybridize with any preexisting $m_J=0$ MZM in odd lobes (not shown), since now all $m_J$ sectors are coupled.

Figure~\ref{fig:MMPD} shows the resulting phase diagrams for both the tubular-core and the solid-core models with disorder, for both the $n=1$ and $n=2$ lobes. Note the remarkable transformation from the corresponding cylindrical cases, Figs.~\ref{fig:TCMPD} and ~\ref{fig:SCMPD}. The topological regions exhibit a characteristic checkerboard-like pattern due to the even-odd effect of the many MZMs, which now occupy roughly half of parameter space. In addition, for the special case of the modified hollow-core model, the invariant $\mathcal{Q}$ can be obtained analytically\cite{Vaitiekenas:S20}, and the topological phase boundaries are given in terms of Eqs.~\eqref{LutchynV}-\eqref{LutchynC} (replacing $R$ with $R_0$, the $\varphi$-averaged $R_{\rm av}$) by the simple algebraic equation
\begin{equation}
\left(\mu_{m_J} - C_{m_J}\right)^2 - \left(A_{m_J} + V^\phi_Z \right)^2 + \Delta^2 = 0.
\end{equation}
Note that each $m_J$ produces a different boundary that is independent of the mode-mixing strength.

The gap opening in the $n=1$ LP lobe and the appearance of a Majorana ZEPs in the $n=2$ LP lobe were also observed in Ref.~\onlinecite{Vaitiekenas:S20}, where a fully two-dimensional numerical model was used.
Subsequently, mode mixing perturbations were also analyzed in Ref.~\onlinecite{Penaranda:PRR20} using a phenomenological model. Compared to these, and leaving aside its lower computational cost, our approach has the key advantage of revealing the underlying mechanism for these effects. In regards to the $n=1$ minigap, it establishes a direct connection between the $m_J$ CdGM splitting and the amplitude of the $2m_J$ angular harmonic of the average wavefunction radius inside the core, see Eq.~\eqref{mode-mixing-terms}. This is the reason why non-smooth disorder creates larger minigaps. This also explains why a perfect hexagonal cross section, as shown in Fig.~\ref{fig:Hex}, is unable to open a full minigap in general, since its harmonics come only in multiples of six.

\section{Summary and conclusions}
\label{Sec:conclusions}

In this work we have studied the rich spectral phenomenology of MZMs in full-shell hybrid nanowires. We recovered previous published results but also went beyond in our analysis, systematically covering different models for the semiconductor core and different regions of parameter space, aiming to rationalize the different underlying mechanisms for the observed behaviors. We have focused on the emergence of MZMs both analyzing the LDOS (or $dI/dV$) at the end of a semi-infinite wire as well as the topological phase diagrams for each model. We first consider pristine full-shell nanowires modelled with the cylindrical approximation, and then in the presence of mode-mixing perturbations. We have found that the cylindrical limit is often an excellent approximation, and constitutes the ideal starting point to understand the system's complex phenomenology.

In the presence of a magnetic flux, full-shell hybrid nanowires are dominated by the LP effect of the shell. The LP effect is characterized by a modulation of the superconducting gap across transitions between phases (lobes) with an increasing integer number $n$ of shell fluxoids as an axial magnetic field is increased. Within each even/odd LP lobe, all states in an infinite and cylindrical nanowire can be indexed by the semi-integer/integer generalized angular momentum $m_J$, and the momentum $k_z$ along the wire. It was shown~\cite{Paya:PRB23} that the occupation of different $m_J$ subbands introduces a collection of finite-energy van-Hove singularities in the LDOS. These features appear below the LP-modulated parent gap and disperse with flux depending on $m_J$. They were dubbed CdGM analogs and were studied in Ref.~\onlinecite{Paya:PRB23}.

The van-Hove singularities in different $m_J$ subbands are the result of the proximity effect of the shell, which opens gaps in the normal bands. The gaps of $m_J\neq 0$ sectors open at finite energy at finite flux, so that their zero-energy density may be finite. The $m_J=0$ sector (only present in odd lobes) is special, as it becomes gapped around zero energy. In the presence of a radial SO coupling, this sector can be mapped into an effective Oreg-Lutchyn model~\cite{Oreg:PRL10,Lutchyn:PRL10}, and can therefore undergo a topological transition into a non-trivial phase, with a MZM appearing at the end of the wire. When the rest of occupied $m_J\neq 0$ subbands are ungapped around zero, the $m_J=0$ MZM will be unprotected against generic perturbations, but may still be visible as a pinned ZEP on top of a finite background in tunnel spectroscopy.

We have characterized the flux extension of this ZEP in the $n=1$ LP lobe using different models for the semiconductor core, and found that it depends on the spatial distribution of the $m_J=0$ electron wavefunction across the wire's section. This dependence was analyzed using four models for the semiconductor core, dubbed hollow-core, tubular-core, modified-hollow-core and solid-core, each with different transverse confinement characteristics and numerical complexity. In the hollow-core approximation, where all the charge is located at the core-shell interface, we found that two symmetric ZEPs appear at the edges of the odd lobes. We found an analytical expression for the MZM flux interval, Eq.~\eqref{Lphi}, as a function of intrinsic and geometrical parameters of the wire, and the superconductor-semiconductor coupling. As the charge density spreads from the interface towards the core axis in a tubular-core nanowire of increasing thickness, the right ZEP disappears, whereas the flux interval of the left one grows, eventually covering the whole lobe for a sufficiently thick semiconductor tube. We found that the tubular-core nanowire can be accurately described with what we called a modified hollow-core approximation (for sufficiently thin tubes) at a reduced computational cost. This approximation was based on the realization that, for a tubular-shaped nanowire, the electron wavefunctions of the different populated $m_J$ subbands are very similar and centered around an average radius $R_{\rm{av}}<R$ that depends on the tube thickness. An analytical expression for the MZM flux interval in this case was given in Eq.~\eqref{LphiMHC}. As a conclusion of this part, we see that it is possible to access information about the charge-density distribution of the core by studying the flux interval of the MZMs in tunneling spectroscopy. Equation~\eqref{LphiMHC} could also be useful to extract the value of some intrinsic parameters of the semiconductor wire (if others were known) that are otherwise inaccessible due to the superconductor encapsulation.

The topological phase diagram for ungapped (unprotected) and gapped (protected) $m_J=0$ MZMs was computed along several axes of the parameter space of each model. For the hollow- and tubular-core models, the phase diagram was found to exhibit two disjoint wedge-shaped regions with ungapped MZMs for positive and negative SO coupling $\alpha$. We found a clear asymmetry between the two that favors $m_J=0$ MZMs with $\alpha>0$ (radial SO axis pointing outwards). The regions with gapped MZMs form smaller islands within the $\alpha>0$ wedge region, and are absent for the $\alpha<0$ wedge. Islands with protected MZMs exist with multiple occupied $m_J\neq 0$ modes, but they eventually disappear with sufficiently high occupation. The solid-core model allows higher radial modes to become occupied, each of which introduces its own replica of the wedged-shaped regions. There are topologically protected islands only for the first wedges corresponding to the occupation of the lowest radial mode (in this case both for upper and lower wedges). Having only the first radial mode occupied is possibly a not very realistic scenario in solid-core nanowires, though. As more radial modes are occupied, CdGM analogs typically fill the $n\ge 1$ LP lobes and no topological minigaps develop. We can conclude from this part that, among disorder-free full-shell nanowires, those with tubular cores (i.e. insulator-semiconductor-superconductor concentric heterowires) are the most advantageous to look for protected MZMs. Moreover, as the tubular-core thickness increases, the minimum SO coupling values needed to enter the topological phase get substantially reduced. In this case, we find large topological minigaps in the $n=1$ LP lobe that can extend across the whole flux interval of the left Majorana ZEP.

The above picture, particularly in regards to the finite-energy spectrum features, remains accurate in the presence of mode-mixing perturbations. These represent deviations from cylindrical symmetry, such as polygonal cross sections or disorder. Mode-mixing, however, introduces changes in the spectrum around zero energy. We first analyzed how the preceding results are modified with a hexagonal cross section, which is the most common shape of crystalline nanowires. In the more extreme case of a thin hexagonal tubular-core nanowire, where the core wavefunction may develop a hexagonal shape, the phase diagram suffers some small changes respect to the cylindrical case. Some narrow stripes appear across the wedge-shaped topological regions where MZMs become split, and some new stripes appear outside the wedges with new MZMs. However, the splittings are small and the topological minigaps of the new regions are approximately one order of magnitude smaller than those of the topologically protected islands. For the parameters away from the narrow stripes, the LDOS remains essentially unchanged with respect to the cylindrical case. As the thickness of the tubular core increases, or in the solid-core scenario, wavefunctions become more cylindrical, so that mode-mixing is weakened and the cylindrical approximation becomes essentially exact.

In the presence of generic mode-mixing perturbations, we found that the changes in the spectrum around zero energy are more dramatic. At a basic level, mode mixing is able to trivially gap the occupied $m_J\neq 0$ modes by coupling them to their $-m_J$ partners, greatly extending the size of the protected MZM islands. Mode-mixing, moreover, introduces a remarkable new mechanism that completely transforms the topological phase diagram. It allows any $\pm m_J$ pair to also undergo a topological phase transition with an associated MZM, both within even and odd lobes. The reason is that mode mixing was analytically found to act as an effective $p$-wave pairing between $\pm m_J$ pairs. The phase diagram is strongly modified as a result, with ubiquitous MZMs extending across approximately half of parameter space within all lobes, following an even-odd checkerboard pattern reminiscent of multimode, D-class, partial-shell nanowires.

In conclusion, our systematic analysis shows that full-shell hybrid nanowires offer a far richer topological
phenomenology than could be initially expected. The combination of radial SO coupling, fluxoid trapping, radial confinement and mode-mixing combine into a fascinating electronic system with high potential for the development of topological phases and the study of Majorana physics. The models and methods developed here provide a numerically efficient and conceptually transparent approach to its phenomenology and its possible extensions.

All the numerical code used in this manuscript was based on the Quantica.jl package \cite{Quantica:Z21} and is available upon reasonable request to the authors. Simulation data and plotting code for the figures in this article are available at Zenodo \cite{Paya:Repository}. Visualizations were made with the Makie.jl package \cite{Danisch:JOSS21}. Pfaffian calculations employed the algorithm of Ref.~\onlinecite{Wimmer:ATMS12}.

\acknowledgments
We gratefully acknowledge discussions with Saulius Vaitiek\.{e}nas, Charles. M. Marcus, Rui E. Silva and Filip K\v{r}í\v{z}ek. This research was supported by Grants PGC2018-097018-B-I00, PID2021-122769NB-I00, PID2021-125343NB-I00 and PRE2022-101362 funded by MICIU/AEI /10.13039/501100011033, ``ERDF A way of making Europe" and ``ESF+".

\appendix

\section{Model}
\label{Ap:Model}

In this work we follow closely the formalism introduced in Refs.~\onlinecite{Vaitiekenas:S20,Paya:PRB23}. In Ref.~\onlinecite{Paya:PRB23}, the semiconductor SO coupling was ignored for the most part, since it was shown to have in general a small effect on the CdGM analog states. However, the SO coupling is essential for the topological superconducting phase and the creation of MZMs. In the following we present a summary of our model following particularly Ref.~\onlinecite{Paya:PRB23} but including the SO interaction. We moreover introduce a couple of extra subsections not described before, where we explain how to calculate the Majorana localization length and a phenomenological model to include mode-mixing perturbations in the Hamiltonian.

\subsection{The Little-Parks effect of the shell}
\label{Ap:LPeffect}

Let us first describe the effect of the threading flux $\Phi$ on the superconducting shell alone, i.e., the blue region in Fig.~\ref{fig:sketch}(a). Consider a hollow superconducting cylinder along the $\hat{z}$ direction of thickness $d$ and inner radius $R$.
A magnetic field $\vec B = B \hat{z}$ is applied along its axis. In the symmetric gauge, the vector potential reads $\vec{A}=\frac{1}{2}(\vec{B}\times\vec{r})=(-y, x, 0)B_z/2 = A_\varphi\hat{\varphi}$, where $A_\varphi = B r/2$. Here $r$ is the radial coordinate and $\varphi$ denotes the azimuthal angle around $\hat{z}$. The magnetic field threads a flux through the cylinder, defined as
\beqa
\label{flux}
\Phi &=& \pi R_\mathrm{LP}^2 B_z,\\
R_\mathrm{LP} &=&R + d/2.\nonumber
\eeqa
Note that $\Phi$ is taken at the mean radius $R_\mathrm{LP}$ of the shell.

In the presence of a threading magnetic flux, a thin superconductor cylinder develops the so-called LP effect \cite{Little:PRL62, Parks:PR64}. This effect is due to the doubly-connected geometry of the superconductor in combination with the magnetic field, which create a quantized winding of the superconductor order parameter phase around the vortex
\beq
\label{Delta}
\Delta(\vec{r}) = \Delta(r)e^{in\varphi}.
\eeq
Here $\Delta(r)=|\Delta(\vec{r})|$ denotes the pairing amplitude. Note that we ignore any $\varphi$ or $z$ dependence of this quantity. The winding number $n$ is an integer, also known as \textit{fluxoid} number~\cite{London:50,Tinkham:04,De-Gennes:18}, that increases in jumps as $\Phi$ grows continuously. Winding jumps are accompanied by a repeated suppression and recovery of the superconducting gap, forming LP \textit{lobes} associated with each $n$.

The shells we consider here can be approximated as dirty superconductors, as also done in previous works \cite{Vaitiekenas:S20,Ibabe:A22}. This approximation is reasonable since carriers in experimental shells experience substantial scattering from the typical oxidation layer that develops on the outer surface \cite{Stanescu:PRB17}, domain walls, impurities and even inhomogeneous strains. Since we are considering a thin superconductor shell, $d~\ll~\lambda_{\rm L}$, where $\lambda_{\rm L}$ is the London penetration length, we can approximate the pairing amplitude to a position-independent constant, $\Delta(r)=\Delta$.

The problem of an ordinary diffusive superconductor in the presence of an external magnetic field is very similar to the
problem of a superconductor containing paramagnetic impurities~\cite{Maki:PTP64,Groff:PR68}. This was originally described by Abrikosov and Gor'kov~\cite{Abrikosov:SPJ61,Skalski:PR64} whose theory was later applied to the LP effect{\cite{Lopatin:PRL05,Shah:PRB07,Dao:PRB09,Schwiete:PRB10,Sternfeld:PRL11}. All the details of these theories can be found in App. A of Ref.~\onlinecite{Paya:PRB23}. Defining $\Lambda$ as a pair breaking parameter,
Abrikosov-Gor'kov found a closed form solution for the pairing amplitude
\beqa
\ln\frac{\Delta(\Lambda)}{\Delta(0)} &=& -P\left(\frac{\Lambda}{\Delta(\Lambda)}\right),\nonumber\\
P(z\leq 1) &=& \frac{\pi}{4}z,\nonumber\\
P(z\geq 1) &=& \ln\left(z+\sqrt{z^2-1}\right)+\frac{z}{2}\arctan\frac{1}{\sqrt{z^2-1}}\nonumber\\
&&-\frac{\sqrt{z^2-1}}{2z},
\label{LP1}
\eeqa
where $\Delta(0)\equiv \Delta_0$ is the pairing of a ballistic superconductor, i.e., for $\Lambda=0$. Note that $\Lambda$ has energy units and is bounded by $0\leq \Lambda\leq \Delta_0/2$. The equation for $\Delta(\Lambda)$ has to be solved self-consistently.

Subsequently, Skalski \textit{et al.}~\cite{Skalski:PR64} found an analytical expression for the energy gap given by
\beqa
\Omega(\Lambda) =  \left(\Delta(\Lambda)^{2/3}-\Lambda^{2/3}\right)^{3/2}.
\label{LP2}
\eeqa
Note that the energy gap $\Omega$ is only equal to the pairing amplitude $\Delta$ in the absence of depairing effects, and is smaller otherwise.

Assuming cylindrical symmetry, a standard Ginzburg-Landau theory of the LP effect \cite{Lopatin:PRL05,Shah:PRB07,Dao:PRB09,Schwiete:PRB10,Sternfeld:PRL11} provides an explicit connection between flux and depairing
\beqa
\Lambda(\Phi) &=& \frac{k_{\rm B} T_{\rm c}\,\xi_{\rm d}^2}{\pi R_{\rm{LP}}^2}\left[4\left(n-\frac{\Phi}{\Phi_0}\right)^2 + \frac{d^2}{R_{\rm{LP}}^2}\left(\frac{\Phi^2}{\Phi_0^2} + \frac{n^2}{3}\right)\right],\nonumber\\
n(\Phi) &=& \lfloor \Phi/\Phi_0\rceil = 0, \pm 1,\pm 2, \dots
\label{LP3}
\eeqa
where $\Phi_0=h/2e$ is the superconducting flux quantum, $\xi_{\rm d}$ is the diffusive superconducting coherence length and $T_{\rm c}$ is the zero-flux critical temperature. At zero field $\Lambda(0)=0$, $\Omega(0) = \Delta(0)\equiv \Delta_0$ and $k_{\rm B} T_{\rm c}\approx\Delta_0/1.76$, where $k_{\rm B}$ is the Boltzmann constant.

The solution for Eqs.~\eqref{LP1}-\eqref{LP3} is qualitatively different depending on the ratios $R_{\rm LP}/\xi_{\rm d}$ and $d/R_{\rm LP}$. It ranges from the non-destructive regime analyzed in the main text ($\Omega$ is always nonzero, satisfied for $R_{\rm LP}/\xi_{\rm d}\gtrsim 0.6$ if $d\rightarrow 0$) to the destructive regime ($\Omega$ vanishes in a finite window around odd half-integer $\Phi/\Phi_0$, satisfied for smaller $R_{\rm LP}/\xi_{\rm d}$), see App.~\ref{Ap:destructive}.

\subsection{The Hamiltonian}
\label{Ap:Hamiltonian}

We now consider the hybrid structure consisting of the superconductor shell and the semiconductor core. In the Nambu basis $\Psi=(\psi_\uparrow, \psi_\downarrow, \psi_\downarrow^\dagger, -\psi_\uparrow^\dagger)$, the Bogoliubov-de Gennes (BdG) Hamiltonian is given by
\begin{eqnarray}
H_{\rm{BdG}}=
\begin{bmatrix}
H_0(\vec{A}) & \Delta(\vec{r})\\
\Delta^*(\vec{r}) & -\sigma_y H_0^*(\vec{A})\sigma_y
\end{bmatrix}.
\label{HBdG}
\end{eqnarray}
Here, $H_0(\vec{A})$ is the Hamiltonian of the hybrid wire in the normal state and $\Delta(\vec{r})$ is the superconducting order parameter in the shell, both in the presence of the magnetic field $\vec{B}=B\hat{z}$ applied along the wire's axis. $\sigma_i$, with $i=(x,y,z)$, are Pauli matrices in the spin sector.

For the core we consider a semiconductor with a large Rashba SO coupling $\alpha$ (such as InAs) owing to the local inversion symmetry breaking in the radial direction at the superconductor-semiconductor interface. The SO coupling is thus proportional to (minus) the electric field that arises at the interface due to the spatially-varying electrostatic potential energy $U(r)$, see Fig.~\ref{fig:sketch}(b). Using a standard approximation from the 8-band model~\cite{Winkler:03}, we can write
\begin{eqnarray}
\label{Eq:SOC}
\alpha(r) &=& -\alpha_0\partial_r U(r), \\
\alpha_0 &=& \frac{P^2}{3}\left[\frac{1}{\Delta_{\rm g}^2}-\frac{1}{(\Delta_\mathrm{soff}+\Delta_{\rm g})^2}\right].\nonumber
\end{eqnarray}
Using the Kane parameter $P = 919.7~\mathrm{meV}\, \mathrm{nm}$, the semiconductor gap $\Delta_{\rm g} = 417$~meV and split-off gap $\Delta_{\rm soff}=390$~meV, relevant for InAs, one obtains $\alpha_0 = 1.19~\textrm{nm}^2$. There are however more elaborate approximations where this value is increased due to confinement effects, see for instance Ref.~\onlinecite{Escribano:PRR20}. It is also possible to have other contributions to the SO coupling, such as the presence of strain at the superconductor-semiconductor interface due to the lattice mismatch between both materials. Therefore, we will consider in the following $\alpha_0$ as a free parameter (that can be both positive and negative) to study the phenomenology of the phase diagram and the MZMs with SO coupling.

We can thus write the low-energy Hamiltonian for the semiconducting core as
\beqa
H_{\rm{core}}&=&\frac{(\vec{p}+eA_\varphi\hat{\varphi})^2}{2m^*}\sigma_0+U\sigma_0-\mu\sigma_0+V_{\rm Z}\sigma_z\nonumber\\
&&+\alpha\vec{r}\cdot[\vec{\sigma}\times(\vec{p}+eA_\varphi\hat{\varphi})],
\eeqa
where $\vec{p}$ is the electron momentum operator, $\mu$ is the semiconductor chemical potential, $m^*$ is the semiconductor effective mass, $e>0$ is the unitary charge and $\sigma_i$ the spin Pauli matrices, with $\sigma_0 = \mathbb{I}$. Even though it is not necessary for the appearance of the topological phase, we also consider the Zeeman effect produced by the magnetic field,
\beq
V_{\rm Z}=\frac{1}{2}g\mu_{\rm B}B_z,
\label{Zeeman}
\eeq
where $\mu_{\rm B}$ is the Bohr magneton and $g$ is the semiconductor Land\'e $g$-factor.

Concerning the non-homogeneous electrostatic potential $U(r)$ inside the core, this potential is a consequence of the band-bending imposed by the epitaxial core/shell Ohmic contact, which in turn stems from the difference of the Al work function and the InAs electron affinity~\cite{Mikkelsen:PRX18, Liu:PRB21, Chen:A23}. We note that the degree of band-bending and precise shape of $U(r)$ depends on the microscopic details of the interface and the self-consistent electrostatic screening. In keeping with our conceptual approach up to this point, we consider a simple model for $U(r)$ of the form
\beq
\label{pot}
U(r) = U_\mathrm{min} + (U_\mathrm{max}-U_\mathrm{min})\left(\frac{r}{R}\right)^2,
\eeq
see Fig.~\ref{fig:sketch}(b).

The normal Hamiltonian $H_0$ in Eq.~\eqref{HBdG} is composed of $H_{\rm{core}}$ and the shell Hamiltonian in the normal state. Since the shell is a dense metal (with much smaller Fermi wavelength than the semiconductor), it is in general quite demanding to include it explicitly in the numerical solution of the Hamiltonian. We then choose to write an \textit{effective} BdG Hamiltonian $H$ of the proximitized nanowire by integrating out the shell degrees of freedom. This procedure introduces a self energy $\Sigma_\mathrm{shell}$ into the Green's function $G(\omega)=\left[\omega - H_\mathrm{core} - \Sigma_\mathrm{shell}(\omega)\right]^{-1}$. This $\Sigma_\mathrm{shell}$ acts on the core surface $r=R$. We thus define the effective BdG Hamiltonian for the system as $H\equiv\omega-G^{-1}(\omega)=H_{\rm{core}}+\Sigma_\mathrm{shell}(\omega)$, which is in general frequency dependent. It can be written as
\begin{eqnarray}
\label{solid}
H &=& \left[\frac{(p_\varphi+eA_\varphi(r) \tau_z)^2 + p_r^2 + p_z^2}{2m^*}+U(r) - \mu\right]\sigma_0\tau_z \nonumber \\
&&+V_{\rm Z}\sigma_z\tau_0+\alpha p_z(-\sin(\varphi)\sigma_x+\cos(\varphi)\sigma_y)\tau_z \nonumber \\
&&-\alpha(p_{\varphi}+eA_{\varphi}(r)\tau_z)\sigma_z\tau_z +\Sigma_\mathrm{shell}(\omega,\varphi),
\end{eqnarray}
where $p_r^2 = -\frac{1}{r}\partial_r(r\partial_r)$, $p_\varphi = -\frac{1}{r}i\partial_\varphi$, $p_z = -i\partial_z$ are the momentum operators for electrons in cylindrical coordinates and $\tau_i$ are Pauli matrices for the electron-hole degree of freedom, with $\tau_0 = \mathbb{I}$. Note that we use $\hbar = 1$ throughout, so that $\omega$ has units of energy.

In the expression above, and in general in the rest of this work, we neglect non-local self-energy components (a valid approximation for disordered shells~\cite{Vaitiekenas:S20}) and also any non-uniformity of the self energy along the wire length, so that $\Sigma_\mathrm{shell}$ depends only on frequency and the angle $\varphi$ around the cylinder axis, $\Sigma_\mathrm{shell}(\omega,\varphi)$.
The form of $\Sigma_\mathrm{shell}$ for a diffusive shell is expressed in terms of a decay rate $\Gamma_{\textrm{S}}$ from the core into the shell (in the normal state),
\begin{equation}
\Sigma_\mathrm{shell}(\omega,\varphi)=\Gamma_{\textrm{S}}\sigma_0\frac{\cos(n\varphi)\tau_x+\sin(n\varphi)\tau_y-u(\omega)\tau_0}{\sqrt{1-u(\omega)^2}}.
\end{equation}
Here, the complex function $u(\omega)$ is obtained as the solution of
\beqa
u(\omega)=\frac{\omega}{\Delta(\Lambda)}+\frac{\Lambda}{\Delta(\Lambda)}\frac{u(\omega)}{\sqrt{1-u(\omega)^2}}.
\label{Eq:u}
\eeqa
Note that $u(\omega)$ depends on the flux $\Phi$ and the fluxoid number $n$ through Eq.~\eqref{LP3}. This equation can be rewritten as a fourth-order polynomial with root $u(\omega)$. We choose the solution that leads to the adequate continuity and asymptotic behavior of the retarded Green's functions. As a consequence, $u(\omega\rightarrow 0) \rightarrow 0$ inside the LP lobes.

\subsection{Quantum numbers}
\label{Ap:quantumnumbers}

The effective BdG Hamiltonian exhibits two symmetries that can be used to classify its eigenstates \cite{Vaitiekenas:S20}.
First, in the limit of infinite wire, the translation symmetry along $z$ leads to a good $k_z$ quantum number. Second, the Hamiltonian exhibits cylindrical symmetry. In the presence of the SO interaction and the pairing winding, $\Delta=\Delta(\varphi)$, the Hamiltonian becomes $\varphi$-dependent, which does not commute with the orbital angular momentum $l_z = -i\partial_\varphi$. However, we can define a generalized angular momentum as  $J_z = -i\partial_\varphi+\frac{1}{2}\sigma_z +\frac{1}{2}n\tau_z$, which is the sum of the orbital angular momentum $l_z$, the spin momentum $s_z=\frac{1}{2}\sigma_z$ and the ``fluxoid momentum" $f_z=\frac{1}{2}n\tau_z$, all of them projected along the $z$ direction. $J_z$ does commute with $H$, $[J_z, H]=0$, so that the eigenvalues $m_J=m_l+m_s+m_n$ of $J_z$ are good quantum numbers of the eigenstates of $H$. Since $m_l\in \mathbb{Z}$ and $m_s\in\pm 1/2$, the possible eigenvalues $m_J$ are
\beq
m_J = \left\{\begin{array}{ll}
\mathbb{Z}+ \frac{1}{2} & \textrm{if $n$ is even} \\
\mathbb{Z} & \textrm{if $n$ is odd}
\end{array}\right.,
\label{mL}
\eeq
which points to qualitative differences between the spectrum in even and odd LP lobes. The canonical transformation $\mathcal{U} = e^{-i(m_J-\frac{1}{2}\sigma_z -\frac{1}{2}n\tau_z)\varphi - ik_z z}$ then reduces $H$ to a $\varphi$-independent $4\times 4$ effective Hamiltonian $\tilde H= \mathcal{U}H\mathcal{U}^\dagger$, where
\beqa
\label{solidrot}
\tilde H &=& \left[\frac{k_z^2+p_r^2}{2m^*}+ U(r)-\mu\right]\sigma_0\tau_z+V_{\rm Z}\sigma_z\tau_0\nonumber\\
&&+\frac{1}{2m^*r^2}\left(m_J-\frac{1}{2}\sigma_z -\frac{1}{2}n\tau_z+\frac{1}{2}\frac{\Phi}{\Phi_0}\frac{r^2}{R_{\rm{LP}}^2} \tau_z\right)^2\sigma_0\tau_z\nonumber\\
&&-\frac{\alpha}{r}\left(m_J-\frac{1}{2}\sigma_z -\frac{1}{2}n\tau_z+\frac{1}{2}\frac{\Phi}{\Phi_0}\frac{r^2}{R_{\rm{LP}}^2}\tau_z\right)\sigma_z\tau_z\nonumber\\
&&+\alpha k_z\sigma_y\tau_z+\Sigma_\mathrm{shell}(\omega, 0).
\eeqa
The self energy has here the simpler expression
\beq
\label{shelfenergy}
\Sigma_\mathrm{shell}(\omega,0)=\Gamma_{\textrm{S}}\sigma_0\frac{\tau_x-u(\omega)\tau_0}{\sqrt{1-u(\omega)^2}},
\eeq
see App.~B of Ref.~\onlinecite{Paya:PRB23}. Note that, inside a LP lobe, $u(\omega\rightarrow~0)\rightarrow~0$ and thus
\beq
\label{Eq:SEomega0}
\Sigma_\mathrm{shell}(0,0)=\Gamma_{\textrm{S}}\sigma_0\tau_x.
\eeq
This $\omega$-independent expression is correct when there is a tunnel coupling between the superconductor and the semiconductor.

The eigenstates $\tilde\Psi_{m_J,k_z}(r)$ of $\tilde H$ are related to the original eigenstates $\Psi_{m_J,k_z}(r,\varphi, z)$ of $H$ by $\Psi_{m_J,k_z}(r,\varphi, z) = \mathcal{U}^\dagger(\varphi, z)\tilde\Psi_{m_J,k_z}(r)$.

\subsection{Numerical methods for Green functions}
\label{Ap:Green}

All the observables of interest (local density of states, differential conductance, Majorana localization length, etc.) are computed in terms of the Green function $g^r$ within the first unit cell of a discretized version of a semi-infinite nanowire, see App.~\ref{Ap:Obs}. The nanowire is described by the rotated BdG Hamiltonian $\tilde{H}(m_J, \omega)$ in Eq.~\eqref{solidrot}, which includes the shell self energy $\Sigma_\mathrm{shell}(\omega)$. In this Appendix we describe how we go from the continuous, differential operator $\tilde{H}$ to a discretized tight-binding version, and how we use it to compute $g^r$.

The conventional approach of discretizing a differential operator $\tilde{H}$ by replacing derivatives by finite differences requires some care in cases where non-Cartesian coordinates are used, like here. If we discretize the $z$ coordinate into discrete sites at fixed $r$ with lattice constant $a_0$, the $z$ derivatives, such as $p_z^2/2m^* = -t^2a_0^2\partial_z^2$ [where $t_0 = 1/(2m^*a_0^2)$], can be trivially transformed into an onsite energy $2t_0$ and a hopping $-t_0$ between neighboring sites. The radial kinetic energy, however, requires taking care of the Jacobian $J=r$ of polar coordinates. We follow the DLL-FDM scheme of Ref.~\onlinecite{Arsoski:CPC15}. The radial coordinate $r$ is also discretized uniformly with a lattice spacing $a_0$ (equal to the $z$ lattice spacing), replacing derivatives with finite differences in the differential eigenvalue equation $\tilde H\psi(r)=\varepsilon\tilde\Psi(r)$. We then absorb the Jacobian $J=r$ of the cylindrical coordinates into modified discrete eigenstates $F(r_i) = \tilde\Psi(r_i)\sqrt{r_i}$ and into the corresponding Hamiltonian $H'= r^{1/2}\tilde H r^{-1/2}$. With this we arrive at a discrete eigenvalue problem $\sum_{i'}{H'}_{ii'} F(r_{i'}) = \varepsilon F(r_i)$ with a Hermitian Hamiltonian matrix ${H'}_{ii'}$, whose discrete eigenstates are, by virtue of their definition, trivially orthonormal without $J$, $\sum_iF^*_\alpha(r_i)F_\beta(r_i)=\delta_{\alpha\beta}$. The kinetic energy  $\tau_z p_r^2/m$ in $H'$ transforms, in the discrete ${H'}_{ii'}$, into an onsite term $o_i=2t_0\tau_z$ plus a radial hopping $t_{ii'} = -t_0\tau_z r/\sqrt{r_ir_{i'}}$ between the nearest neighbors, where $t_0=1/(2m^*a_0^2)$. Note that the $r/\sqrt{r_ir_{i'}}$ factor directly stems from the cylindrical Jacobian, but does not break the symmetry $t_{ii'}=t_{i'i}$. Also, when applying the above DLL-FDM scheme to systems including the origin $r=0$, the correct boundary condition must be implemented there. This is done by excluding the $r=0$ site and multiplying $o_i$ at the $r=a_0$ site by 3/4~\cite{Arsoski:CPC15}.

The above procedure yields a discretization of the nanowire in terms of individual unit cells, each with $N$ radially distributed sites at a fixed $z=n a_0$. The intra-cell Hamiltonians (which includes the shell self energy at the core-shell boundary, and hence depends on $\omega$) is an $4N\times 4N$ matrix (the 4 comes from spin and electron-hole degrees of freedom), denoted by $h$. The inter-cell Hamiltonians are dubbed $h_+$ (hop towards positive $z$) and $h_- = (h_+)^+$ (hop towards negative $z$). All of these depend on $m_J$.

The retarded Green function at the first unit cell is denoted by $g^r$, and is an $4N\times 4N$ matrix that satisfies the Dyson equation
\beqa
\label{dyson}
g^r &=& g_0 + g_0\Sigma^r g_r = (g_0^{-1} - \Sigma^r)^{-1},\\
g_0 &=& (\omega-h)^{-1},\\
\Sigma^r &=& h_- g^r h_+,
\eeqa
or equivalently
\beq
\label{dyson2}
h_+ - g_0^{-1}(g^rh_+) + h_-(g^rh_+)^2 = 0.
\eeq

To solve Eq.~\eqref{dyson2} we consider a diagonalization of the $g^rh_+$ operator, $g^rh_+ = \phi\lambda\phi^{-1}$, where $\phi$ is the eigenvector matrix and $\lambda$ the diagonal (complex) eigenvalue matrix. The above equation then becomes a quadratic eigenvalue equation
\beq
\label{lambda}
\left(h_+ - g_0^{-1}\lambda + h_-\lambda^2\right)\phi = 0.
\eeq
The eigenmodes $\phi$ decay as $\phi(n) = \lambda^n\phi$ as we move $n$ unit cells away from the end of the nanowire. If we add an small imaginary part to $\omega \to \omega+i0$, any eigenvalue $\lambda$ with $|\lambda|<1$ will correspond to a retarded mode, either a bound or a causally propagating state.
The type of quadratic eigenvalue equation \ref{lambda} can be solved by linearlizing it with the auxiliary matrix $\chi = \phi\lambda$, so that Eq.~\eqref{lambda} becomes
\beq
\label{pencil}
\Bigg[\overbrace{\left(\begin{array}{cc}
0 & 1\\ -h_+ & g_0^{-1}
\end{array}\right)}^{A}
-
\lambda\overbrace{\left(\begin{array}{cc}
1 &0 \\ 0 &h_-
\end{array}\right)}^{B}\Bigg]
\overbrace{\left(\begin{array}{c}
\phi \\ \chi
\end{array}\right)}^{\psi} = (A-\lambda B)\psi = 0.
\eeq
The solutions to this equation can be obtained using standard linear algebra algorithms. To compute $g^r$, we first compute $\Sigma_r = h_-g^rh_+ = h_-\phi\lambda\phi^{-1}$ from the retarded solutions, and then use $g^r = (g_0^{-1} - \Sigma^r)^{-1}$.

\subsection{LDOS, $dI/dV$ and Majorana localization length}
\label{Ap:Obs}

The LDOS at the end of a semi-infinite hollow-core nanowire is given by
\beq
\rho(\omega) = -\frac{1}{\pi}\sum_{m_J}\mathrm{Im}\,\mathrm{Tr}\, g^r_{m_J}(\omega).
\eeq
Here $g^r_{m_J}(\omega)$ is the retarded Green function at the end of the nanowire, computed as in App.~\ref{Ap:Green}. The trace $\mathrm{Tr}$ is taken over the radial sites and electron/hole degree of freedom.

For the computation of the $dI/dV$ we couple the semi-infinite nanowire to a normal lead across a tunnel barrier. The lead has a  similar Hamiltonian as the nanowire, but without a shell and with a higher carrier density. The lead and barrier introduce a retarded self energy $\Sigma^r_{\rm L}$ into the nanowire that can be computed as $\Sigma^r_{\rm L} = h_+ g^r_{m_J,\mathrm{L}} h_-$, where $g^r_{m_J,\mathrm{L}}$ is the surface Green function of the decoupled lead plus barrier. The differential conductance is obtained from the Blonder-Tinkham-Klapwijk (BTK) formula \cite{Blonder:PRB82}
\beq
dI/dV = \frac{e^2}{h}\left[N_p-\mathrm{Tr}(r_{ee}^+r_{ee})+\mathrm{Tr}(r_{he}^+r_{he})\right],
\eeq
where $N_p$ is the number of propagating modes in the normal lead, $r_{ee}$ is the normal reflection matrix from the lead modes onto the barrier, and $r_{he}$ is the Andreev reflection matrix. This expression can be recast in terms of the retarded $G^r = \left(\left(g^r\right)^{-1} - \Sigma^r_{\rm L}\right)^{-1}$ and advanced $G^a=(G^r)^+$ Green functions of the coupled nanowire at the contact, and the decay rate matrix $\Gamma = i\left(\Sigma^r_{\rm L} - \left({\Sigma^r_{\rm L}}\right)^+\right) $ from the nanowire into the lead as
\beqa
dI/dV &=&\frac{e^2}{h}\left\{ i\mathrm{Tr}\left[(G_{ee}^r-G_{ee}^a)\Gamma_{ee}\right] - \mathrm{Tr}\left[G_{ee}^a\Gamma_{ee} G_{ee}^r\Gamma_{ee}\right]\right.\nonumber\\
&&\left.+\mathrm{Tr}\left[G_{eh}^a\Gamma_{hh} G_{he}^r\Gamma_{ee}\right]\right\}.
\eeqa

Finally, the Majorana decay length is computed from the retarded $\lambda$ eigenvalues obtained in App.~\ref{Ap:Green} when computing $g^r$ of the decoupled lead.
The eigenmodes $\phi$ decay as $\phi(n) = \lambda^n\phi$ as we move $n$ unit cells away from the end of the nanowire. Hence, any eigenvalue $\lambda$ with $|\lambda|<1$ will correspond to a decaying bound state concentrated to the end of the nanowire. If we fix $\omega=0$ and $m_J=0$, this is a Majorana bound state, and its decay length $\xi_{\rm M}$ is related to $\lambda$ by $|\lambda_{\rm max}| = \exp(-a_0/\xi_{\rm M})$, where $\lambda_{\rm max}$ is the retarded eigenvalue that has the largest modulus, and $a_0$ is the lattice constant of the discretized nanowire (see App.~\ref{Ap:Green}). Note that, as a consistency check, the presence of a Majorana implies a gaped $m_J=0$ sector, so that all $|\lambda|$ should remain smaller than 1, even as $\mathrm{Im}\left\{\omega\right\} \to 0$, except precisely at the topological transition where the gap closes.

\subsection{Mode-mixing perturbations}
\label{Ap:modemixing}

We wish to introduce mode-mixing perturbations in our model to analyze the robustness of $m_J=0$ MZMs in the presence of gapless $m_J\neq 0$ modes. Modeling arbitrary cross-section deformations or cross-section disorder in an exact way requires diagonalizing a Hamiltonian that, apart from the periodic $z$-coordinate, depends on two Cartesian coordinates $(x,y)$, and possibly averaging over disorder realizations. This can be computationally expensive, specially for large cross-section areas. In this section we devise an extension of the modified hollow-core model that can help us understand the consequences of mode mixing perturbations at a greatly reduced computational cost.

The model starts from the cylindrical approximation we have used so far in all our models. The cylindrical approximation is very convenient because it removes one coordinate, $\varphi$, in the resolution of the Hamiltonian. This is achieved by transforming the Hamiltonian of Eq.~\eqref{solid} into a basis with a good $m_J$ quantum number, Eq.~\eqref{solidrot}. In this way, apart from $z$, the resolution of the Hamiltonian for each fixed $m_J$ only requires diagonalizing the $r$ sector. If the nanowire under study is such that the charge density is not too far away from the superconductor-semiconductor interface, we can moreover use the modified hollow-core approximation, as explained in Sec.~\ref{Sec:TCNw}. In this case we may fix $r$ to a certain average radius $R_{\rm av}$ and thus the Hamiltonian only depends on $z$. The transverse wavefunction has circular symmetry and it is characterized by a constant radius $R_{\rm av}(\varphi)=R_0$.

Now, to take into account mode mixing in a computationally efficient way, we introduce angular perturbations in the average radius $R_{\rm av}$ as a Fourier series of the form
\begin{equation}
    \label{harmonics}
    R_{\rm av}(\varphi) = R_0 + \mathrm{Re}\sum_{\ell=1}^{\ell_\mathrm{max}} \delta R_\ell \exp\left(i\ell \varphi\right),
\end{equation}
where $\ell>0$ and Re means real part.

To describe a hexagonal cross section we can compute the $\delta R_\ell$ coefficients as the Fourier series of the following hexagonal distortion
\begin{equation}
    \label{hexagon}
    R_\mathrm{av}(\varphi) = R_0 \frac{c_0}{\cos{\varphi} + \frac{1}{\sqrt{3}}\sin{\varphi}},
\end{equation}
where $c_0 = \pi/[3\sqrt{3}\,\mathrm{arctanh}(1/2)]\approx 1.0066$ is taken so that the $\varphi$-averaged radius is equal to $R_0$. The first three non-zero coefficients in Eq.~\eqref{harmonics} are then $\delta R_6 = 0.0581563 R_0$, $\delta R_{12} = 0.0161251 R_0$, $\delta R_{18} = 0.00733844 R_0$. These harmonics yield an increasingly accurate expansion of a regular hexagon. We may instead wish to model a $R_\mathrm{av}(\varphi)$ with the shape of a smooth hexagon (more similar to a circle), e.g. to capture the kind of smoother angular distribution of the electron cloud in a hexagonal nanowire as the average radius moves closer to the nanowire axis. In this case we can simply scale down these few $\delta R_\ell$ by a factor between zero and one.

To instead describe random geometric distortions of the cross section, the $\delta R_\ell$ coefficients are taken as independent, complex, random variables. The modulus $|\delta R_\ell|$ is then modeled to follow a Gaussian distribution with zero mean and standard deviation $\sigma_\ell = \sigma_1/\ell^2$ for a fixed $\sigma_1$, and the random phase $\mathrm{arg}\left(\delta R_\ell\right)$ is taken uniformly distributed in the interval $[0,2\pi]$. The resulting $R_{\rm av}(\varphi)$ is a randomly distorted cross section, with the peculiarity that the variance of the distortion $\sigma^2 = \langle \left[R_{\rm av}(\varphi)-R_0\right]^2\rangle$ and the derivative $\langle \left[\partial_\varphi R_{\rm av}(\varphi)\right]^2\rangle$ both remain bounded as the $\ell_\mathrm{max}$ cutoff is increased. The strength of the resulting random distortion is then controlled by a single dimensionless scale $\sigma/R_0$, where $\sigma^2 = \frac{1}{2}\sum_\ell\sigma_\ell^2 =\frac{\sigma_1^2}{2}\sum_\ell \ell^{-4} = (\pi^4/180)\sigma_1^2$.

Finally, we can also describe a stronger random inter-mode mixing model similar to the above, but with $\sigma_\ell = \sigma_1/\ell$. The variance of the distortion in this model is still bounded as the cutoff is increased, but not its derivative. This type of model could represent atomic-sized defects, such as core-shell interface dislocations or amorphous Aluminum oxide on the the shell surface. In this case, the variance of the distortion is $\sigma^2 = \frac{\sigma_1^2}{2}\sum_\ell \ell^{-2} = (\pi^2/12)\sigma_1^2$.

In the presence of an $R_{\rm av}(\varphi)$ perturbation, we cannot simply replace $r =R_{\rm av}$ in the solid-core Hamiltonian and ignore the radial momentum as we did to obtain the modified hollow-core model in Sec.~\ref{Sec:TCNw}. Let us assume though that the wavefunction can still be decomposed in generalized angular momentum $m_J$ modes of the same $J_z$ operator defined previously, so that we can project our Hamiltonian onto a set of wavefunctions sharply localized at $R_{\rm av}(\varphi)$ for each $\varphi$,
\begin{equation}
    \bra{\vec{r}}\ket{m_J}  = \frac{1}{\sqrt{2\pi}} \frac{\delta_{\sigma_r}(r - R_{\rm av}(\varphi))}{\sqrt{r \delta_{\sigma_r}(0) / \sqrt{2}}} e^{i (m_J - \frac{1}{2}\sigma_z - \frac{1}{2}n\tau_z ) \varphi} \tilde{\Psi}_{m_J}.
\end{equation}
Here $\tilde{\Psi}_{m_J}$ is the Nambu spinor and $\delta_{\sigma_r}(r)$ is a quasi-Dirac delta function defined as a Gaussian centered at $r=0$ with a small standard deviation $\sigma_r\ll R_0$,
\begin{equation}
    \delta_{\sigma_r}(r) = \frac{1}{\sqrt{2\pi}\sigma_r} e^{-\frac{r^2}{2\sigma_r^2}}.
\end{equation}
This way, the diagonal matrix blocks $\mel{m_J}{H}{m_J}$ remain as in Sec.~\ref{Sec:HCNw}, but additional non-diagonal, $m_J$-mixing blocks appear. Due to the shape of the wavefunction, only two terms of the Hamiltonian of Eq.~\eqref{solid} contribute to the mode mixing,
\begin{align}
   M_1(r, \varphi) &= (p_\varphi + e A_\varphi (r) \tau_z)^2 \tau_z,  \label{kinetic}\\
    M_2 (r, \varphi, z) &= -\alpha p_\varphi \sigma_z \tau_z - \alpha e A_\varphi (r) \sigma_z \tau_z.  \label{Calpha}
\end{align}

Evaluating for our harmonically deformed radial profiles (Eq.~\eqref{harmonics}) and defining $\delta R_{-\ell} = \delta R_{\ell}^*$ for all $\ell$, the non-diagonal matrix elements become
\begin{widetext}
\begin{eqnarray}
    \mel{m_J}{M_1(r, \varphi)}{m_J \pm \ell} &=& \frac{\delta R_{\pm\ell}}{R_0^3} \left[ \left( m_J \pm \frac{\ell}{2} - \frac{1}{2}\sigma_z - \frac{1}{2}n \tau_z \right)^2 + \frac{\ell^2}{4} - \left( e A_\varphi (R_0) \right)^2 R_0^2 \right]\tau_z, \nonumber\\
    \mel{m_J}{M_2 (r, \varphi, z)}{m_J \pm \ell} &=& \alpha \frac{\delta R_{\pm\ell}}{2R_0^2} \left(m_J \pm \frac{\ell}{2} - \frac{1}{2}\sigma_z - \frac{1}{2} n \tau_z - R_0 e A_\varphi (R_0) \tau_z \right)\sigma_z \tau_z.
\label{mode-mixing-terms}
\end{eqnarray}
\end{widetext}
We also have to take in account that the area enclosed by this wavefunction is
\begin{equation}
    \int_0^{2\pi}d\varphi\int_0^{R_\mathrm{av}(\varphi)} r dr = \left[ 1 + \frac{1}{2} \sum_\ell \left(\frac{\delta R_\ell}{R_0} \right)^2 \right] \pi R_0^2,
\end{equation}
and thus the vector potential is related to the flux as
\begin{equation}
    A_\varphi (R_0) = \frac{1}{2 + \sum_\ell{\left( \frac{\delta R_\ell}{R_0} \right)^2}} \frac{R_0}{R_{\rm LP}^2} \frac{\Phi}{\Phi_0}.
\end{equation}
Note that the $\varphi$-dependent wavefunction shift is therefore only included through its effect on the kinetic and SO terms, which include the coupling to the magnetic flux. We neglect here the effect of the shift on the strength of the induced pairing, in $\Sigma_\mathrm{shell}$, since this dependence is subject to a greater uncertainty in microscopic modeling, and should not invalidate the main kinetic effect discussed here.

\section{Topological phase in the hollow-core approximation}
\label{Ap:HCA_Topology}

\begin{figure}
   \centering
   \includegraphics[width=\columnwidth]{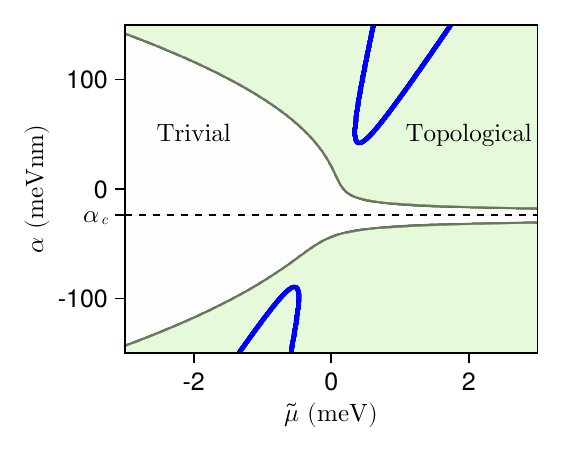}
   \caption{\textbf{Trivial and topological phases for the hollow-core model.} Trivial (white) and topological (green) phases as a function of SO coupling $\alpha$ and Fermi energy $\tilde\mu$ for a hollow-core nanowire with $R=70$~nm, $d=0$, $\Gamma_{\rm S}=\Delta_0$ and $g=0$. The green region shows the parameters for which $\tilde\mu>\tilde\mu_{\rm c}$, i.e., for which there can be topological phase transitions for some values of $\Phi$. For $\alpha=\alpha_{\rm c}=-1/(2m^*R)$ the system is always trivial. The blue contour shows the topological phase diagram, i.e., the parameter regions for which there actually are Majorana ZEPs in the first lobe. Note that the upper blue contour is the same as Fig.~\ref{fig:HCA}(g). Other parameters as in Fig.~\ref{fig:HCA}.}
   \label{fig:HCATopology}
\end{figure}

We have shown in the main text that, in the hollow core-model, it is possible to get an analytical expression for the flux at which topological transitions occur, Eq.~\eqref{PhiTT}. For convenience, let us rewrite this equation as
\begin{equation} \label{eq:phiAB}
    \frac{\Phi_{\rm TT}^{(i)}}{\Phi_0} = 1 \pm \sqrt{A \pm B},
\end{equation}
where
\begin{align}
    A &\equiv 1 + 4 m^* R (\alpha + 2 m^* R \alpha^2 + 2 R \tilde\mu ), \label{eq:A} \\
    B &\equiv 4 R \sqrt{m^* C}, \label{eq:B} \\
    C &\equiv (m^* \alpha^2 + 2 \tilde\mu )(1 + 2 m^* R \alpha)^2 - 4 m^* R^2 \Gamma_{\rm S}^2, \label{eq:C}
\end{align}
where $R$ is the hollow-core radius, $m^*$ the electron effective mass, $\alpha$ the SO coupling, $\tilde\mu$ the Fermi energy and $\Gamma_{\rm S}$ the decay rate from the core into the shell. Note that we neglect the Zeeman effect by taking $g=0$. If $g\neq 0$, it is not possible to get an analytical expression for $\Phi_{\rm TT}^{(i)}$.

If $C > 0$ ($B \in \mathbb{R}$) we have four real solutions (all combinations of $\pm$): two inner ones ($i=2$ taking the $-/-$ and $i=3$ the $+/-$ signs) and two outer ones, typically outside of the first lobe ($i=1$ taking the $-/+$ and $i=4$ the $+/+$ signs). As $m^*, R, \Gamma_{\rm S} \geq 0$ by definition, it can be shown that $A \geq B$ is always true. For $A = B$, two solutions are degenerate at $\Phi = \Phi_0$, but this only happens if $\Gamma_{\rm S} \rightarrow 0$.

If $C < 0$ ($B \notin \mathbb{R}$) we have four complex solutions. Therefore, the system does not present any topological phase transition for any magnetic flux.

In terms of the system parameters, the boundary between these two situations is given by a critical Fermi energy
\begin{equation}
\tilde\mu_{\rm c} =\frac{2 m^* R^2 \Gamma_{\rm S}^2}{(1 + 2 m^* R \alpha)^2} - \frac{m^* \alpha^2}{2},
\end{equation}
i.e., Eq.~\eqref{muc} of the main text. The topological phase of the wire is then only possible if $\tilde\mu > \tilde\mu_{\rm c}$, see the green region of Fig.~\ref{fig:HCATopology} as a function of $\alpha$ and $\tilde\mu$. Note that $\tilde\mu\rightarrow\infty$ for $\alpha\rightarrow\alpha_{\rm c}$, where
\begin{equation}
\alpha_{\rm c}=-\frac{1}{2m^*R}.
\label{alphac}
\end{equation}
In the vicinity of $\alpha_{\rm c}$ the wire cannot be topological for any Fermi energy, see white region in Fig.~\ref{fig:HCATopology}. This divides the topological phase in two disjoint regions, one with $\alpha>\alpha_{\rm c}$ and another with $\alpha<\alpha_{\rm c}$.

The value of $\alpha_{\rm c}$ can be understood in the following way. Following Ref.~\onlinecite{Vaitiekenas:S20}, we can rewrite the effective BdG Hamiltonian \eqref{solidrot} in the hollow-core approximation (and taking $g=0$) as
\begin{multline}
    \tilde{H}_{\rm HC} = \left( \frac{p_z^2}{2 m^*} - \tilde\mu_{m_J}\right)\tau_z + V_{\rm Z}^{\phi} \sigma_z + A_{m_J}
    \\ + C_{m_J} \sigma_z \tau_z + \alpha p_z \sigma_y \tau_z + \Sigma_{\text{shell}}(\omega, 0),
\end{multline}
where
\begin{eqnarray}
    \label{LutchynV}
    \tilde\mu_{m_J} &=&  \tilde\mu - \frac{\alpha}{2 R} - \frac{1}{8 m^* R^2} \left( 4 m_J^2 + 1 + \phi^2 \right), \\
    V_{\rm Z}^{\phi} &=& \frac{1}{2}\phi \left( \frac{1}{2 m^* R^2} + \frac{\alpha}{R} \right),
    \label{LutchynZeeman}\\
    A_{m_J} &=& -\frac{m_J\phi}{2 m^* R^2},  \\
    C_{m_J} &=& -m_J \left(\frac{1}{2 m^* R^2} + \frac{\alpha}{R}\right), \label{LutchynC}
\end{eqnarray}
with $\phi=n-\Phi(R)/\Phi_0$. For $m_J=0$, $A_{m_J}=C_{m_J}=0$ and it is then possible to map this Hamiltonian to the conventional 1D Majorana nanowire model \cite{Lutchyn:PRL10,Oreg:PRL10}. Note that the \textit{effective} Zeeman term $V_{\rm Z}^{\phi}$ has an orbital origin here, and it is present even in the absence of semiconductor $g$ factor. Written in this way, it is clear that $\alpha$ has three different effects on the bands of the system for $m_J=0$. One is to shift the effective Fermi energy $\tilde\mu_{m_J}$, another is through the standard SO term $\alpha p_z \sigma_y$, and the third one is to modify  $V_{\rm Z}^{\phi}$.

The presence of the effective Zeeman field $V_{\rm Z}^{\phi}$ is required for the topological phase transition. Interestingly, $V_{\rm Z}^{\phi}\rightarrow 0$ as $\alpha\rightarrow\alpha_{\rm c}$. The SO coupling is thus capable of canceling the effective Zeeman energy at precisely $\alpha_{\rm c}$, explaining the asymptotes in Fig.~\ref{fig:HCATopology}(a). Moreover, $\alpha_{\rm c}$ separates two topological regions for which the effective Zeeman field is positive ($\alpha>\alpha_{\rm c}$) and negative ($\alpha<\alpha_{\rm c}$). A way to understand the value of $\alpha_{\rm c}$ is by noticing that a particle living in a circle of radius $R$, threaded by a flux $\Phi$, has an orbital kinetic energy $E_{p_{\varphi}}=(m_l+\frac{1}{2}\Phi/\Phi_0)^2/(2m^*R^2)$, where $m_l$ is its orbital angular momentum, see App.~\ref{Ap:quantumnumbers}. In the $m_J=0$ sector we have a spin-up $m_l=1$ electron, and spin-down $m_l=0$ electron, which are therefore split by an orbital kinetic energy $E_{p_{\varphi}}(m_l=1) - E_{p_{\varphi}}(m_l=0) = (\Phi/\Phi_0-1)/(2mR^2)$. In the presence of a radial SO coupling, these two states acquire an additional SO energy shift $E_{\alpha}=\sigma_z(m_l+\frac{1}{2}\Phi/\Phi_0)\alpha/R$, so that their total splitting becomes the $V_{\rm Z}^{\phi}$ in Eq.~\eqref{LutchynZeeman} above. Hence, given the relative orientation of spin and $m_l$ in the $m_J=0$ sector, an SO coupling pointing radially outwards (inwards) will add (subtract) to the orbital splitting.

Finally, notice that the green regions in Fig.~\ref{fig:HCATopology} correspond to parameters for which there can be real solutions of $\Phi_{\rm}^{(i)}$, Eq.~\eqref{PhiTT}. But these flux values can be outside of the first LP lobe, so that the system is not topological. The topological phase diagram, defined as the regions for which the MZM flux interval of Eq.~\eqref{Lphi} is $L_{\Phi}>0$, is contoured by solid blue curves in Fig.~\ref{fig:HCATopology}. This is a much smaller region than the green one. Note that these topological boundaries are symmetric with respect to the point $\alpha=\alpha_{\rm c}$ and $\tilde\mu=-\alpha/2R$. Since $\alpha_{\rm c}$ is a negative quantity, the $\alpha>0$ topological region starts at smaller values of SO coupling than the $\alpha<0$ one in absolute value.

\begin{figure}
   \centering
   \includegraphics[width=\columnwidth]{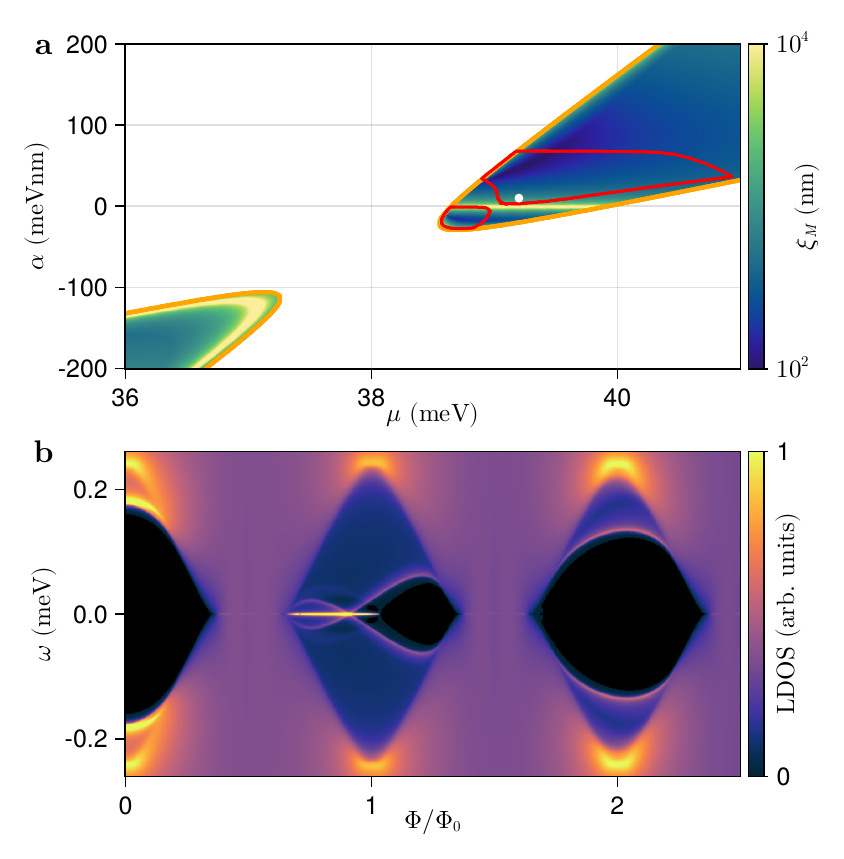}
   \caption{\textbf{Tubular-core model in the destructive Little-Parks regime.}
(a) Topological phase diagram as a function of constant $\alpha$ and $\mu$ for a tubular-core nanowire with $R = 30$~nm, $d=0$ and $W = 10$~nm. Parameter islands with topologically protected MZMs are bounded by red lines. The solid orange curve has been calculated using the modified hollow-core approximation, with $R_{\rm av}=24.5$~nm, $\tilde{\mu} \in [-2.38, 2.62]$ and $\Gamma_{\rm S}^{\rm av} = 1.9 \Delta_0$. (b) LDOS (in arbitrary units) as a function of $\omega$ and $\Phi/\Phi_0$ for the parameters of the red dot in (a): $\alpha=10$~meV\,nm and $\mu=39.1$~meV. $\Gamma_{\rm S}=8\Delta_0$ chosen so that the degeneracy points in the $n=0$ LP lobe remain around $\omega = \pm 0.18$~meV. There is a topological minigap $E_{\rm g}=20$~$\mu$eV for $\Phi=0.99\Phi_0$. Other parameters as in Fig.~\ref{fig:HCA}.}
\label{fig:TCMdestructive}
\end{figure}

\section{Results in the destructive Little-Parks regime}
\label{Ap:destructive}

In the main text we have studied a representative case in the non-destructive LP regime (with $R=70$~nm, $d\sim 10$~nm, $\xi_{\rm d}=70$~nm and thus $R_{\rm LP}/\xi_{\rm d}\gtrsim 0.6$, see App.~\ref{Ap:LPeffect}). Here we show results for a full-shell nanowire in the destructive LP regime.

Particularly, we consider a tubular-core model with $R=30$~nm, $d=0$~nm and $W=10$~nm. The topological phase diagram as a function of constant SO coupling $\alpha$ and chemical potential $\mu$ can be seen in Fig.~\ref{fig:TCMdestructive}(a). In Fig.~\ref{fig:TCMdestructive}(b) we show the LDOS as a function of energy $\omega$ and normalized flux $\Phi/\Phi_0$ for the wire parameters corresponding to the red dot in Fig.~\ref{fig:TCMdestructive}(a), i.e., $\alpha=10$~meV\,nm and $\mu=39.1$~meV.
The qualitative behavior of the LDOS is similar to that of Fig.~\ref{fig:TCMpos}. However, there are a couple of important differences. Now the induced LP gap $\Omega(\Phi)$ closes in between lobes, creating normal regions around half-integer $\Phi_0$ values. These normal regions contribute to diminish the flux windows of the different LP lobes, and bends the GdGM analogs towards zero energy at the lobe edges more pronouncedly than in the non-destructive LP case.

On the other hand, the values of $\alpha_{\rm min}$ for which the system can enter into the topological phase decrease considerably with respect to nanowires with larger radius. This was shown already in the topological phase diagram of Fig.~\ref{fig:HCA}(i) and it is confirmed in Fig.~\ref{fig:TCMdestructive}(a), where $\alpha_{\rm min}\approx 0$. This means that we can have MZMs with very small and realistic values of SO coupling.  In this case, for $\alpha=10$~meV\,nm we get a maximum topological minigap $E_{\rm g}=20$~$\mu$eV for $\Phi\approx\Phi_0$, but larger topologically-protected MZM windows and minigaps can be obtained by slightly increasing $|\alpha|$.

\section{Mode mixing as an effective $p$-wave pairing}
\label{Ap:pwave}

The fact that mode-mixing can cause a topological band inversion is non-trivial. The easiest way to understand it is by establishing a mapping between the effect of mode-mixing and a $p$-wave pairing in a spinless one-dimensional conductor. Consider the paradigmatic spinless superconductor with $p$-wave pairing, with Bogoliubov Hamiltonian
\begin{equation}
H(k_z)=\left(\begin{array}{cc}
\frac{k_z^2}{2m}-\mu &  k_z \Delta  \\
k_z \Delta   & -\frac{k_z^2}{2m}+\mu
\end{array}\right)
\end{equation}
If $\mu=\Delta=0$ the spectrum consists of two parabolic bands with dispersion $\pm \frac{k_z^2}{2m}$ touching at $k=\epsilon = 0$. As $\Delta$ becomes finite at $\mu=0$, the parabolic touching point becomes a linear crossing around $k_z = 0$, signaling a topological phase transition between $\mu < 0$ (trivial) and $\mu>0$ (non-trivial, with MZM at the ends of the superconductor). The topological invariant $\mathcal{Q}$ is given by the sign of the Pfaffian of the antihermitian matrix $W(k_z) = \tau_x H(k_z)$ at $k_z=0$,
\begin{equation}
    \mathcal{Q} = \mathrm{sign}~\mathrm{Pf}[W(0)],
\end{equation}
with $\mathcal{Q}=1$ signaling a trivial phase and $\mathcal{Q}-1$ a non-trivial one.
For our spinful BdG Hamiltonian of Eq.~\eqref{HBdG} one should define $W(k_z) = \tau_y\sigma_y H(k_z)$ instead~\cite{Ghosh:PRB10}. The Pfaffian may be efficiently computed numerically using the algorithm of Ref.~\onlinecite{Wimmer:ATMS12}.

Consider now the four modes in a $m_J\neq 0$ sector of the full-shell nanowire. This $m_J$ may be an integer (odd lobes) or a half-integer (even lobes). These modes have approximately parabolic dispersion $\epsilon_{m_J}^{(\nu)}(k_z)$ ($\nu=\pm 1, \pm 2$) around $k_z=0$, with eigenstates $|k_z,m_J,\nu\rangle$. For finite flux $\Phi$ they may cross zero energy. Crucially the dispersion of the $-m_J$ sector is just the opposite $\epsilon_{-m_J}^{(-\nu)}(k_z) = -\epsilon_{m_J}^{(\nu)}(k_z)$. Singling out one of these four subband pairs (e.g., $\nu=1$) we have again two touching parabolas at a specific value of $\Phi$ and $\mu$ where the mode becomes populated/unpopulated. At this special point, a mode mixing perturbation like Eqs.~\eqref{mode-mixing-terms} could be expected to lift the $k_z=0$ degeneracy. Remarkably, however, it can be shown analytically that

\begin{equation}
    \langle k_z,-m_J,-\nu|M_{1}|k_z,m_J,\nu\rangle \propto \alpha k_z,
\end{equation}

\begin{equation}
    \langle k_z,-m_J,-\nu|M_{2}|k_z,m_J,\nu\rangle \propto  \alpha^2 k_z.
\end{equation}

We omit the full expression, since it is quite involved and is not required for our discussion. Consequently, we find that mode mixing acts exactly as the $p$-wave pairing term in the spinless superconductor when acting on the two conjugate $\epsilon_{\pm m_J}^{\pm \nu}$ eigenstates. Therefore, mode mixing transforms the parabola crossing into a topological band inversion, and hence introduces a MZM for each of these inversions. The total invariant is a product of the corresponding $\mathcal{Q}$, as long as no special selection rules exist that forbids mode mixing between specific $m_J$ (the case of pristine hexagonal section nanowires is an exception to this). As a result, any odd occupation of the $m_J$ modes with generic mode mixing will result in a nontrivial $\mathcal{Q}=-1$ (including even lobes), while an even occupation will be trivial with $\mathcal{Q}=-1$. The total number of Majoranas will be $N_{\rm M} = 1$ if $\mathcal{Q}=-1$, and $N_{\rm M}=0$ otherwise.

If two sectors are not coupled by mode mixing (like $m_J=\pm 3$ and $m_J=0$ in the hexagonal case) they can both contribute with independent MZMs, and then $N_{\rm M}$ may be greater than 1.

\bibliography{biblio_reduced}

\end{document}